\documentclass[12pt]{article}

\usepackage{cite}
\usepackage{color}
\usepackage{xr}
\usepackage{amsmath,amsfonts,amssymb}
\usepackage{graphicx}
\usepackage[affil-it,auth-lg]{authblk}
\usepackage{setspace}
\usepackage[inner=1.25in,outer=1in,top=1in,bottom=1.35in,foot=0.8in,dvips]{geometry}

\newcommand*{\vect}[1]{\boldsymbol{#1}}

\newcommand{\mb}[1]{\mathbf{#1}}

\author[1]{Pavel I. Zhuravlev}
\author[2]{Michael Hinczewski}
\author[1]{Shaon Chakrabarti}
\author[3]{Susan Marqusee}
\author[1]{D.Thirumalai}

\affil[1]{Biophysics Program, Institute for Physical Science and Technology, Department of Chemistry \& Biochemistry, University of Maryland, College Park, MD, 20742}
\affil[2]{Department of Physics, Case Western Reserve University, Cleveland, OH, 44106}
\affil[3]{California Institute for Quantitative Biosciences, Department of Molecular and Cell Biology, University of California, Berkeley, CA, 94720}

\title{Force-dependent switch in protein unfolding pathways and transition state movements}
\begin{document}

\maketitle

\begin{abstract}
 Although known that single domain proteins fold and unfold by parallel pathways, demonstration of this expectation has been difficult to establish in experiments.  Unfolding rate, $k_\mathrm{u}(f)$, as a function of force $f$, obtained in single molecule pulling experiments on src SH3 domain, exhibits upward curvature on a $\log k_\mathrm{u}(f)$ plot. Similar observations were reported for other proteins for the unfolding rate $k_\mathrm{u}([C])$. These findings imply unfolding in these single domain proteins involves a switch in the pathway as $f$ or $[C]$ is increased from a low to a high value. We provide a unified theory demonstrating that if $\log k_\mathrm{u}$ as a function of a perturbation ($f$ or $[C]$) exhibits upward curvature then the underlying energy landscape must be strongly multidimensional. Using molecular simulations we provide a structural basis for the switch in the pathways and dramatic shifts in the transition state ensemble (TSE) in src SH3 domain as $f$ is increased.  We show that a single point mutation shifts the upward curvature in $\log k_\mathrm{u}(f)$ to a lower force, thus establishing the malleability of the underlying folding landscape. Our theory, applicable to any perturbation that affects the free energy of the protein linearly, readily explains movement in the TSE in a $\beta$-sandwich (I27) protein and single chain monellin as the denaturant concentration is varied. We predict that in the force range accessible in laser optical tweezer experiments there should be a switch in the unfolding pathways in I27 or its mutants.
\end{abstract}


\textbf{Abbreviations:} SOP-SC, self-organized polymer with side-chains; WMD, weakly multidimensional; SMD, strongly multidimensional; SMFS, single molecule force spectroscopy; AFM, atom force microscopy; LOT, laser optical trapping

\textbf{Significance Statement:} Single domain proteins with symmetrical arrangement of secondary structural elements in the native state are expected to fold by diverse pathways. However, understanding the origins of pathway diversity, and the experimental signatures for identifying it, are major challenges, especially for small proteins with no obvious symmetry in the folded states. We show rigorously that upward curvature in the logarithm of unfolding rates as a function of force (or denaturants) implies that the folding occurs by diverse pathways. The theoretical concepts are illustrated using simulations of src-SH3 domain, which explain the emergence of parallel pathways in single molecule pulling experiments and provide structural description of the routes to the unfolded state.  We make testable predictions illustrating the generality of the theory.

\section*{Introduction}
	
That single domain proteins should fold by parallel or multiple pathways emerges naturally from theoretical considerations\cite{Harrison:1985wo,Wolynes:1995p2172,Bryngelson1995} and computational studies\cite{Guo:1992dd,Leopold:1992te,Klimov:2005eb,Noe:2009en}. The generality of the conclusions reached in the theoretical studies is sufficiently compelling, which makes it surprising that they are not routinely demonstrated in typical ensemble folding experiments. The reasons for the difficulties in directly observing parallel folding or unfolding pathways in monomeric proteins can be appreciated based on the following arguments. Consider a protein that reaches the folded state by two different pathways. The ratio of flux through these pathways is proportional to $\exp\left[\frac{\Delta G_L^{\ddagger} - \Delta G_H^{\ddagger}}{k_BT}\right]$, where $\Delta G_L^{\ddagger}$ and $\Delta G_H^{\ddagger}$ are, respectively, the free energy barriers separating the folded and unfolded states along the two pathways, $k_B$ is the Boltzmann constant, and $T$ is the temperature.  If $\Delta \Delta G^{\ddagger} = \Delta G_H^{\ddagger} - \Delta G_L^{\ddagger}>0$ is large compared to $k_BT$ then the experimental detection of flux through the high free energy barrier pathway ($H$) is unlikely.  External perturbations (such as mechanical force $f$ or denaturants $[C]$) could reduce $\Delta \Delta G^{\ddagger}$. However, the values of $f$ (or $[C]$) should fall in an experimentally accessible range for detecting a potential switch between pathways. Despite these inherent limitations, Clarke and coworkers, showed in a most insightful study that unfolding of immunoglobulin domain (I27) induced by denaturants occurs by parallel routes\cite{Wright:2003cs}. Subsequently, additional experiments on single chain monellin \cite{Aghera:2012dp}, using denaturants and spectroscopic probes, have firmly shown the existence of multiple paths. Thus, it appears that multiple folding routes can be detected in standard folding experiments \cite{Sosnick:2011ej,Lindberg:2007ga} provided the flux through the higher free energy barriers is not so small that it escapes detection. In addition,  parallel folding pathways have been observed in repeat proteins, where inherent symmetry in the connectivity of the individual domains\cite{Aksel:2014kn} results in parallel assembly.

Single molecule pulling experiments in which $f$ is applied to specific locations on the protein have demonstrated that unfolding of many proteins follow complex multiple routes. Mechanical force, unlike denaturants, does not alter the effective microscopic interactions between the residues, thus allowing for a cleaner interpretation. More importantly, by following the fate of many individual proteins the underlying heterogeneity in the routes explored by the protein can be revealed.  Indeed, pulling experiments and simulations on a variety of single domain proteins\cite{Mickler:2007ky,Stigler:2011ct,Kotamarthi:2013jv} show clear signatures of many routes for $f$-induced unfolding. It could be argued that in many of these studies the network of states connecting the folded and unfolded states is a consequence of complex topology, although they are all single domain proteins. However, the src SH3 domain is a small protein with no apparent symmetry in the arrangement of secondary structure elements which folds in an apparent two-state manner. Thus, the discovery that it unfolds using parallel pathways\cite{Jagannathan:uw,Guinn:2015gq} is unexpected and requires a firm theoretical explanation and structural interpretation.

In single molecule pulling experiments, performed at constant force or constant loading rate, only a one dimensional coordinate, the molecular extension ($x$), is readily measurable. When performed at constant $f$, it is possible to generate folding trajectories ($x$ as a function of time), from which  equilibrium one-dimensional free energy profiles, $F(x)$, can be extracted using rigorous theory\cite{Hinczewski:2013kd}. The utility of $F(x)$ hinges on the crucial assumption  that all other degrees of freedom in the system including the solvent come to equilibrium on time scales much faster than in $x$, so that $x$ itself may be considered to be an accurate reaction coordinate.

A straightforward way to assess if a one-dimensional picture is adequate, is to analyze the $f$ dependence of the unfolding rate $k_\mathrm{u}(f)$, which can be experimentally obtained at constant $f$ or computed from unfolding rates measured at different loading rates\cite{Hyeon07JPHysCondMat,Hinczewski:2013kd}. The observed upward curvature in the [$\log k_\mathrm{u}(f),f]$ plot in src SH3 \cite{Jagannathan:uw}, was shown to be a consequence of unfolding by two pathways, one dominant at low forces and the other at high forces. It was succinctly argued that the measured [$\log k_\mathrm{u}(f),f]$ data cannot be explained by multiple barriers in a one dimensional (1D) $F(x)$ or a 1D profile with a single barrier in which the unfolding rate is usually fit using the  Bell model $k\mathrm{u}(f)=k^0\exp\frac{fx^{\ddagger}}{kT}$, where $k^0$ is the unfolding rate at $f=0$, and $x^{\ddagger}$ is the location of the barrier in $F(x)$ at zero force with respect to the folded state. (Throughout this paper, by location of the barrier, or the transition state, we mean the location with respect to the folded state.) The upward curvatures in the monotonic [$\log k_\mathrm{u}(f),f]$  as well as [$\log k_\mathrm{u}([C]),[C]]$ plots, observed experimentally, necessarily imply that parallel routes are involved in the unfolding process. (A non-monotonic $[\log k_\mathrm{u}(f),f]$ plot suggests catch bond behavior).

In order to provide a general framework for a quantitative explanation of a broad class of experiments, we first present a rigorous theoretical proof that upward curvature in [$\log k_\mathrm{u}(f),f]$ (or [$\log k_\mathrm{u}([C]),[C]]$) implies that the folding landscape is strongly multidimensional (SMD). Hence, such SMD landscapes cannot be reduced to 1D or  superposition of physically meaningful 1D landscapes, which can rationalize the observed convex [$\log k_\mathrm{u} (f),f$] plot. We note {\it en passant} that the shape of the measured [$\log k_\mathrm{u}(f),f]$ plot cannot be justified using $F(x)$ even if $x^{\ddagger}$ were allowed to depend on $f$, moving towards the folded state as $f$ increases. The theory only hinges on the assumption that the perturbation ($f$ or $[C]$) is linearly coupled to the effective energy function of the protein. To illustrate the structural origin of the upward curvature in the [$\log k_\mathrm{u}(f),f]$ plot we also performed simulations of $f$-induced unfolding of src SH3 domain, in order to identify the structural details of the unfolding pathways including the movement of transition states as the force is increased. The results of the simulations semi-quantitatively reproduce the experimental [$\log k_\mathrm{u}(f),f$] plots for both the wild-type (WT) and the V61A mutant. More importantly, we also provide the structural basis of the switch in the unfolding pathways as $f$ is varied, which cannot be obtained using pulling experiments. We obtain structures of the transition state ensembles (TSEs), demonstrating the change in the TSE structures as $f$ is increased from a low to a high value.

\section*{Results}



\subsection*{Nature of the energy landscape from [$\log k_\mathrm{u}(f), f$] plots:} Let us assume that the unfolding rate $k_\text{u}(\alpha)$, as a function of a controllable external perturbation $\alpha$, can be measured. We assume that $\alpha$ decreases the stability of the folded state linearly, as is the case in the pulling experiment with $\alpha=f$, the force applied at two points of the protein. However, the discussion below is quite general, and applies to any external parameter with a linear, additive contribution to the effective protein energy function. For a protein under force, the total free energy has the general form $U(x,\mb{r},f) = U_0(x,\mb{r}) - f x$, with a force contribution $f x$, where $x$ is the end-to-end extension of the protein. Here, $U_0$ is the free energy in the absence of applied tension, and the vector $\mb{r}$ represents all the additional conformational degrees of freedom besides $x$. 


  In the derivation below, we model the
  dynamics of the protein as diffusion of a single particle on the
  multidimensional landscape $U(x,\mb{r},f)$. The unfolding of the
  protein would correspond to the particle starting in the reference
  protein conformation $(x_\text{f},\mb{r}_\text{n})$ in the folded
  state energy basin F and diffusing to any other conformation, with a
  given extension $x_\text{u}> x_\text{f}$, representing the unfolded
  basin U (Fig.~\ref{ws}). The unfolding time for a particular
  trajectory is the time when the particle reaches the target
  conformation for the first time (known as first passage
  time). Averaging this time over all trajectories yields the mean
  first passage time (MFPT) from the unfolded to folded state which we
  denote as $t_\text{u}(x_\text{f},\mb{r}_\text{n},f)$, or the average
  unfolding time. The unfolding rate is the inverse of the unfolding
  time,
  $k_\text{u}(f) \equiv 1/t_\text{u}(x_\text{f},\mb{r}_\text{n},f)$.

  We are interested in finding the curvature of $\log k_\mathrm{u}(f)$
  as a function of $f$, and in particular the sign of
  $\frac{d^2}{df^2}\log k_\text{u}$. Starting from the diffusion
  equation, we find expressions for the MFPT from any
  conformation with extension $x_f$,
  $t_\text{u}(x_\text{f},\mb{r},f)$, and then for
  $\log k_\mathrm{u}(f)$ and its first two derivatives. It turns out
  that if we use the assumptions of a single unfolding pathway, the
  second derivative is negative and the curvature of
  $[\log k_\mathrm{u}(f),f]$ has to be downward.

  The summary of the subsequent derivation is as follows: 1) we start
  from the equation for $t_\text{u}(x_\text{f},\mb{r},f)$ which can be
  obtained from the diffusion equation\cite{Kampen:1992p1933}, 2)
  integrate it over the ${\mb{r}}$ degrees of freedom, 3) use two
  assumptions for evaluating the integral with
  $t_\text{u}(x_\text{f},\mb{r},f)$ inside, 4) solve the ODE
  for the unfolding time, 5) establish that the solution implies certain
  constraints on the shape of the [$\log k_\mathrm{u}(f), f$]
  plot. Following this derivation in detail is not necessary for
  understanding the other parts of the paper.

%
%

  The equation that the MFPT $t_\text{u}(x,\mb{r},f)$ can be
  obtained from the diffusion equation (in Fokker-Planck form) by
  integration over $x$,$ \mb{r}$, and $t$, followed by some
  rearrangements \cite{Kampen:1992p1933}.  The result is called the
  backward Kolmogorov equation:


\begin{equation}\label{d1}
	\begin{split}
D(x,\mb{r}) e^{\beta U(x,\mb{r},f)} \left\{\frac{\partial}{\partial x}\left[ e^{-\beta U(x,\mb{r},f)} \frac{\partial}{\partial x} t_\text{u}(x,\mb{r},f)\right]\right\} + \\ D(x,\mb{r}) e^{\beta U(x,\mb{r},f)} \left\{\nabla_{\mb{r}} \cdot \left[e^{-\beta U(x,\mb{r},f)} \nabla_{\mb{r}} t_\text{u}(x,\mb{r},f)\right]\right\}=-1,
\end{split}
\end{equation}
with the boundary condition $t_\text{u}(x_\text{u},\mb{r},f)=0$, with $\beta=1/k_BT$ and $D(x,\mb{r})$ being the diffusion constant, which for generality is allowed to depend on the conformation. By dividing both sides of Eq.~\eqref{d1} by $D e^{\beta U}$ and integrating over the conformational coordinates $\mb{r}$, we obtain

\begin{equation}\label{d2}
\frac{\partial}{\partial x}\left[ \int d\mb{r}\,e^{-\beta U(x,\mb{r},f)} \frac{\partial}{\partial x} t_\text{u}(x,\mb{r},f)\right] =-\int d\mb{r}\,D^{-1}(x,\mb{r})e^{-\beta U(x,\mb{r})}.
\end{equation}
To get the result in Eq.\ref{d2}, we have assumed that $U(x,\mb{r}) \to \infty$ at the integration limits of the coordinate space of $\mb{r}$, i.e. the diffusion process is bounded. We rewrite Eq.~\eqref{d2} as,

\begin{equation}\label{d3}
\frac{\partial}{\partial x}\left[e^{\beta fx} \int d\mb{r}\,e^{-\beta U_0(x,\mb{r})} \frac{\partial}{\partial x} t_\text{u}(x,\mb{r},f)\right] =-e^{\beta f x} G(x),
\end{equation}
where $G(x) \equiv \int d\mb{r}\,D^{-1}(x,\mb{r})\exp(-\beta U_0(x,\mb{r}))$.

Further simplification of the MFPT expression depends on the nature of the multidimensional free energy $U_0(x,\mb{r})$. In particular, we define a class of free energies that satisfy the following two conditions:
\begin{enumerate}
\item $U_0(x,\mb{r})$ has a single minimum with respect to $\mb{r}$ at
  each point $x$ in the range $x_\text{f}$ to $x_\text{u}$.  We denote
  the location of this minimum as $\mb{r}_\text{m}(x)$.
\item The Boltzmann factor $\exp(-\beta U_0(x,\mb{r}))$ for $\mb{r}$
  near $\mb{r}_\text{m}(x)$ is sharply peaked, so the thermodynamic
  contribution from conformations with coordinates far from
  $\mb{r}_\text{m}(x)$ is negligible.  In other words, we assume fast
  equilibration along the $\mb{r}$ coordinates at each $x$, compared
  to the timescale of first passage between N and U.
\end{enumerate}

A schematic illustration of a $U_0(x,\mb{r})$ satisfying these requirements is shown in Fig.~\ref{ws}A. Diffusion is essentially confined to a single, narrow reaction pathway in the multidimensional space. We will call any $U_0(x,\mb{r})$ in this category weakly multidimensional (WMD) with respect to $x$, since the diffusion process is quasi-1D in terms of the reaction coordinate $x$. In contrast, any $U_0(x,\mb{r})$ that violates either one of the above conditions will be called strongly multidimensional (SMD), since it has characteristics that qualitatively distinguish it from any one-dimensional diffusion process. Note that this categorization makes no other assumptions about the shape of $U_0(x,\mb{r})$ except for those specified above: for example, there could be one or many free energies barriers separating N and U, or none at all. Fig.~\ref{ws}B and C show two examples of $U_0(x,\mb{r})$ that are SMD. In both cases, condition 1 is violated, because in the range $x_\text{f}<x<x_\text{u}$ there is no unique minimum in $U_0(x,\mb{r})$ along $\mb{r}$. For panel B, there are two possible reaction pathways between N and U, while for panel C there is a single pathway, but it is non-monotonic in $x$.

For an energy landscape that is WMD, there are rigorous bounds on the first and
second derivatives of $\log k_\text{u}(f)$ with respect to $f$. To derive
these bounds, note that the WMD assumptions allow us to make a saddle-point
approximation to the integral over $\mb{r}$ on the left-hand side of
Eq.~\eqref{d3}, setting the value of $\mb{r}$ in $\partial
t_\text{u}(x,\mb{r},f)/\partial x$ to $\mb{r}_\text{m}(x)$. Since this will be
the dominant contribution, we approximate Eq.\ref{d3} 

\begin{equation}\label{d4}
\frac{\partial}{\partial x}\left[e^{\beta fx} \frac{\partial}{\partial x} t_\text{u}(x,\mb{r}_\text{m}(x),f) \int d\mb{r}\,e^{-\beta U_0(x,\mb{r})} \right] \approx -e^{\beta f x} G(x).
\end{equation}

By simplifying the notation by defining $\tau_\text{u}(x,f) \equiv t_\text{u}(x,r_\text{m}(x),f)$ and $I(x) \equiv \int d\mb{r}\,\exp(-\beta U_0(x,\mb{r}))$, Eq.~\eqref{d4} becomes

\begin{equation}\label{d5}
\frac{\partial}{\partial x}\left[e^{\beta fx} \frac{\partial}{\partial x} \tau_\text{u}(x,f) I(x) \right] \approx -e^{\beta f x} G(x).
\end{equation}
The solution for $\tau_\text{u}(x,f)$ from Eq.~\eqref{d5}, with boundary condition $\tau_\text{u}(x_\text{u},f)=0$, can be written as a Laplace transform of a function $H(x,s)$,

\begin{equation}\label{d6}
	\begin{split}
\tau_\text{u}(x,f) = \int\limits_0^\infty ds\,e^{-\beta f s} H(x,s),\\
 \qquad H(x,s) \equiv  \int\limits_x^{x_\text{u}} dx^\prime I^{-1}(x^\prime)G(x^\prime-s).
\end{split}
\end{equation}

Both $I(x)$ and $G(x)$ are non-negative functions (since $D(x,\mb{r}) >0$ and $\exp(-\beta U(x,\mb{r})) \ge 0$ for all $x$ and $\mb{r}$), so the function $H(x,s)$ is likewise non-negative, $H(x,s) \ge 0$ for $x\le x_\text{u}$. From this property, it follows that $\tau_\text{u}(x,f) \ge 0$, and

\begin{equation}\label{d7}
\begin{split}
\frac{\partial}{\partial f}\tau_\text{u}(x,f) = -\int\limits_{0}^{\infty} ds\,\beta s\, e^{-\beta fs} H(x,s) \le 0,\\
\frac{\partial^2}{\partial f^2}\tau_\text{u}(x,f) = \int\limits_{0}^{\infty} ds\,(\beta s)^2 e^{-\beta fs} H(x,s) \ge 0,
\end{split}
\end{equation}
for $x \le x_\text{u}$. Since the experimental data is typically plotted in terms of $\log k_\text{u}(f) = - \log \tau_\text{u}(x_\text{f},f)$ with respect to $f$, we are specifically interested in the corresponding derivatives of $\log k_\text{u}(f)$,

\begin{equation}\label{d8}
\frac{d}{df}\log k_\text{u} = -\frac{1}{\tau_\text{u}}\frac{\partial \tau_\text{u}}{\partial f}, \qquad \frac{d^2}{df^2}\log k_\text{u} = \frac{1}{\tau_\text{u}^2}\left[\left(\frac{\partial \tau_\text{u}}{\partial f}\right)^2 - \tau_\text{u} \frac{\partial^2 \tau_\text{u}}{\partial f^2}\right].
\end{equation}

From Eq.~\eqref{d7} we see that $d \log k_\text{u}/d f \ge 0$. The sign of $d^2\log k_\text{u}/df^2$ requires establishing the sign of the term in the square brackets in Eq.~\eqref{d8}, which can be done by using the Cauchy-Schwarz inequality. Let us define two functions, $g_1(x,s) \equiv \Theta(s) \sqrt{e^{-\beta f s} H(x,s)}$ and $g_2(x,s) \equiv \Theta(s) \beta s \sqrt{e^{-\beta f s} H(x,s)}$, where $\Theta(s) =1$ for $s\ge 0$ and 0 for $s<0$. Then from Eqs.~\eqref{d6}-\eqref{d7} we obtain
$\tau_\text{u} = \int\limits_{-\infty}^\infty ds\,|g_1(x,s)|^2$,$\qquad \frac{\partial \tau_\text{u}}{\partial f} = -\int\limits_{-\infty}^\infty ds\, g_1^\ast(x,s) g_2(x,s)$,$\qquad \frac{\partial^2 \tau_\text{u}}{\partial f^2} = \int\limits_{-\infty}^\infty ds\,|g_2(x,s)|^2.$
Using the Cauchy-Schwarz inequality
\begin{equation}\label{d10}
\left|\int\limits_{-\infty}^\infty ds\, g_1^\ast(x,s) g_2(x,s) \right|^2 \le \int\limits_{-\infty}^\infty ds\,|g_1(x,s)|^2 \int\limits_{-\infty}^\infty ds\,|g_2(x,s)|^2
\end{equation} 
we find that $(\partial \tau_\text{u}/\partial f)^2 \le \tau_\text{u} \partial^2 \tau_\text{u}/\partial f^2$. Hence, from Eq.~\eqref{d8} we see that $d^2\log k_\text{u}/df^2 \le 0$. In summary, we can now state the full criterion for the validity of WMD for describing force-induced unfolding:

\subsection*{Criteria for WMD unfolding landscape:} The unfolding
  rate $k_\text{u}(f)$ on a WMD free energy landscape under applied
  force $f$ must satisfy:
\begin{equation}\label{d11}
\frac{d}{df}\log k_\text{u} \ge 0, \qquad \frac{d^2}{df^2}\log k_\text{u} \le 0.
\end{equation}
If $k_\text{u}(f)$ fails to satisfy either of the conditions in Eq.\ref{d11}, the underlying free energy landscape must be strongly multidimensional, and analyses of the measured data using the end-to-end distance, $x$, as a reaction coordinate are incorrect.

\subsection*{Scenarios for [$\log k_\mathrm{u}(f),f$] plots in WMD and SMD:} 
\begin{figure}
	\includegraphics[width=\textwidth]{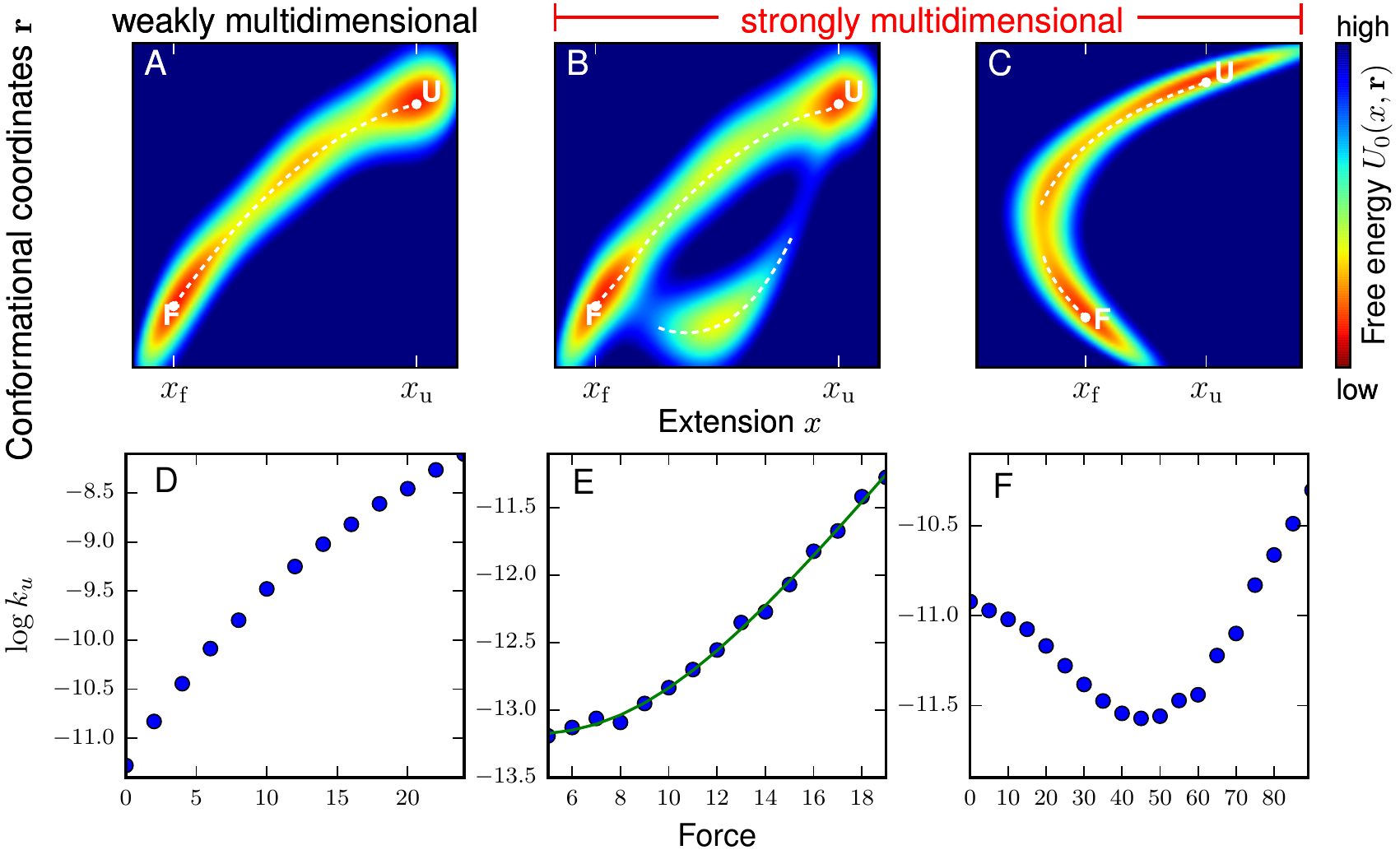}
\caption{Schematic illustration of three multidimensional free
  energy landscapes $U_0(x,\mb{r})$ at zero force, plotted in terms of
  end-to-end extension $x$ versus the other conformational
  degrees of freedom $\mb{r}$.  N and U could represent the native and unfolded
  basins. The white dotted lines are local minima of
  $U_0(x,\mb{r})$ with respect to $\mb{r}$ at a given $x$.  The
  landscape in panel A is weakly multidimensional with respect to $x$,
  according to the conditions described in the main text, while panels
  B and C are strongly multidimensional. (D,E,F) The $[\log k_\mathrm{u}(f), f]$ plots for the energy landscapes given in (A,B,C) obtained from Brownian dynamics simulations. The WMD landscape (left) produces downward curvature. The landscape suggesting parallel pathways (B) produces upward curvature (E), where the solid line is the least squares fit to a double exponential model. The SMD in (C) produces even stronger upward curvature, giving rise to catch bond behavior with $k_\mathrm{u}(f)$ decreasing at low forces and subsequently increasing at larger $f$ values.}\label{ws}
\end{figure}

A range of behaviors in [$\log k_\mathrm{u}(f), f$] plots can be obtained depending on the nature of the energy landscape. Stochastic simulations in a WMD (Figs.\ref{ws}A and D) show that [$\log k_\mathrm{u}(f), f$] has a minor downward curvature, which is readily explained by a generalized Bell model in which the transition state location is allowed to move towards the folded state in accord with the Hammond effect\cite{Hyeon:2006ii}. In contrast, in the SMD (Figs.\ref{ws}B,C and E,F) the [$\log k_\mathrm{u}(f), f$] plot shows upward curvature. The upward curvature in Fig.\ref{ws}E indicates loss of flux from the folded state through two channels in Fig.\ref{ws}B, similar to parallel pathways in protein unfolding experiments. Interestingly, the upward curvature in Fig.\ref{ws}F from the SMD landscape in Fig.\ref{ws}C does not come from parallel pathways. Instead, the lifetime of the folded state first increases followed by the usual decrease as $f$ increases. Such a counterintuitive ``catch bond'' behavior is well documented in a number of protein complexes\cite{Marshall:2003di,Craig14Science,Chakrabarti:2014ke}. The results in Figs. \ref{ws}E and \ref{ws}F show that violations of Eq.\ref{d11} implies that the underlying energy landscape must be SMD.

\subsection*{Naive analyses of the $f$-dependence of $k_\mathrm{u}(f)$:} 
\begin{figure}
	\includegraphics[width=\textwidth]{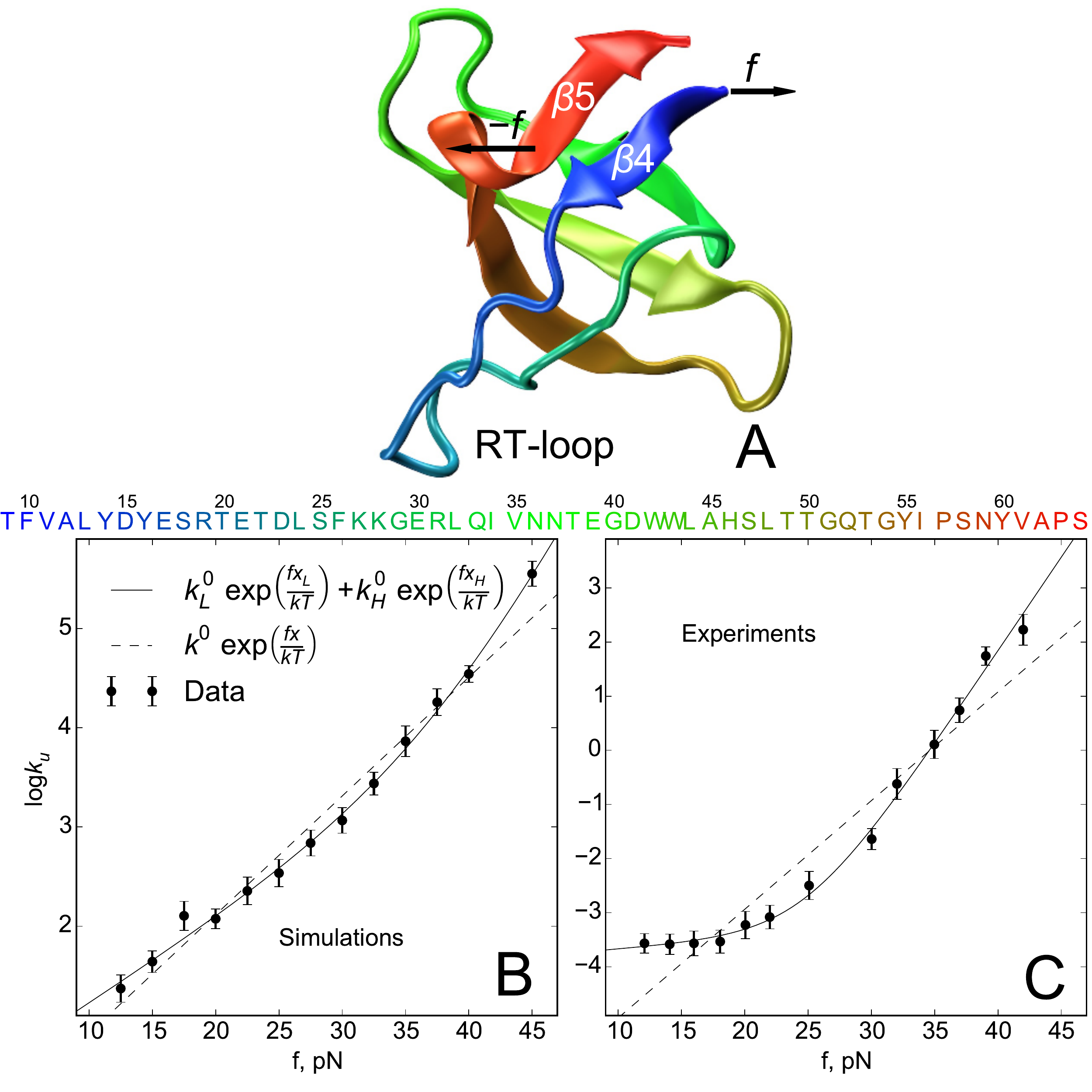}
   	\caption{(A) The native structure of \textit{G.Gallus} src SH3 domain from Tyr kinase, whose sequence is given below. The N-terminal and C-terminal $\beta$-strands are denoted as $\beta 4$ and $\beta 5$. In the simulations, the force is applied to residues 9 and 59. RT-loop is the longest loop of the domain, which is positioned in sequence right after the N-terminal $\beta$-strand; (B) The dependence of unfolding rate $k_\mathrm{u}(f)$ obtained from the simulations using the SOP-SC model as a function of $f$ given in terms of [$\log k_\mathrm{u}(f), f$] plot. The solid line is a two exponential fit (Eq.\ref{eq:2exp}) whereas the dashed line is a fit using Bell model; (C) Same as (B) except the data are obtained from single molecule pulling experiments\cite{Jagannathan:uw}.
\label{fig:twoslopes}}
\end{figure}

Using the data generated by molecular dynamics simulations of force unfolding the src SH3 domain, with force applied to residues 9 and 59, for a set of forces ${f_i}$, 
we calculated $k_\mathrm{u}(f_i)$ for each force $f_i$ as the inverse of the mean first passage time from the folded state to the unfolded state, by averaging the set of first passage times to unfolding ${t_{ij}}$ over the trajectory index $j$ (see Methods). The [$\log k_\mathrm{u} (f),f$] plot for $f$-induced unfolding is non-linear with upward curvature implying that the free energy landscape is SMD (Fig.\ref{fig:twoslopes}B). We note parenthetically that the inadequacy of the Bell model cannot be fixed using movement of the transition state with $f$ or using a one-dimensional free energy profile with two (or more) barriers. Remarkably, the slope change in the simulations qualitatively coincides with measurements on the same protein\cite{Jagannathan:uw,Guinn:2015gq}, where constant force was applied to the residues 7 and 59. Thus, both simulations and experiment show that the condition in Eq.\ref{d11} is violated, implying that the free energy landscape for SH3 is SMD.

The observed $k_\mathrm{u} (f)$ dependence can be fit using a sum of two exponential functions\cite{Jagannathan:uw},
\begin{equation}
 k_\mathrm{u}(f)=k^0_L\exp\frac{f x_L}{kT}+k^0_H\exp\frac{f x_H}{kT}.
 \label{eq:2exp}
\end{equation}
The parameters $k^0_L$ and $k^0_H$ (unfolding rates at $f=0$) and $x_L$ and $x_H$ (putative locations of the transition states) can be precisely extracted using maximum likelihood estimation (MLE, see Methods). According to the Akaike information criterion\cite{Akaike:1974ih}, the double exponential model is significantly more probable than the single-exponential model, for both simulations (relative likelihood of the models $P_2/P_1 \sim 10^3$) and experiments\cite{Guinn:2015gq} ($P_2/P_1 \sim 10^{17}$). The extracted values of $x_L$ and $x_H$, shown in Table \ref{tbl:fitparams}, are $x_L=0.40$ nm, $x_H=1.16$ nm for the simulations data, and $x_L=0.1$ nm, $x_H=1.44$ nm for the experimental data from Ref.\cite{Guinn:2015gq} with the MLE procedure, which differ somewhat from the values reported in Ref.\cite{Guinn:2015gq} ($x_L=0.18$ nm, $x_H=1.52$ nm). Given that the error in $x_L$ estimated for experimental data using MLE is large we surmise that the simulations and experiments are in good agreement. The switch in the forced unfolding behavior (estimated as the point where the third derivative of $\log k_\mathrm{u}(f)$ in Eq.\ref{eq:2exp} with parameters given by MLE changes sign) occurs around 25 pN for the experimental data and around 35 pN for the simulation data. These comparisons show that the simulations based on the self-organized polymer with side-chains(SOP-SC) model reproduce quantitatively the shape of the [$\log k_\mathrm{u}(f)$] plot. Because simulations are done by coarse-graining the degrees of freedom, involving both solvent and proteins, the $k_\mathrm{u}(f)$ from simulations are expected to be larger than the measured values with the discrepancy being greater at higher forces. Our previous work\cite{Liu:2011jc} showed that the unfolding rate in denaturants is larger by a factor of $\approx$ 150, which is similar to the difference between experiment and simulations in Fig.\ref{fig:twoslopes}. However, because the inference about parallel pathways relies solely on the shape of $\log k_\mathrm{u}(f)$ the inability to quantitatively reproduce the precise value of $k_\mathrm{u}(f)$ is irrelevant. 

Despite the good fits to Eq.\ref{eq:2exp} neither
$x_L$ nor $x_H$ can be associated with transition state location as is traditionally assumed. We show below that such projections onto a one dimensional coordinate cease
to have physical meaning when the underlying folding landscape is SMD. The apparent barriers to unfolding at $f=0$ along the pathways
can be estimated using $k^0_L\approx k_\mathrm{u0}\exp{\left(-\Delta G^\ddagger_L/k_BT\right)}$ and $k^0_H\approx k_\mathrm{u0}\exp{\left(-\Delta
G^\ddagger_H/k_BT\right)}$. Using the accepted estimates for the prefactor ($k_\mathrm{u0} \approx 10^{6} - 10^{7} \mathrm{s}^{-1}$)\cite{Li:2004gn,Yang2003,Kubelka:2004bu}, and the values of $k^0_L$ and $k^0_H$ from the fits of experimental data (Table \ref{tbl:fitparams}), we obtain $\Delta
G^\ddagger_L \approx (18 - 20) k_BT$, depending on the value of $k_\mathrm{u0}$ and $\Delta G^\ddagger_H \approx (26 - 28) k_BT$. If these values are reasonable
then the ratio of fluxes through the two pathways at $f=0$ is $\sim \exp(-(\Delta G^\ddagger_H - \Delta G^\ddagger_L)/k_BT) \sim
e^{-8} = 3 \cdot 10^{-4}$, which is much smaller than those obtained in the simulations by direct calculation of the flux through the two pathways.
In addition, the finding that $x_H>x_L$ also makes no physical sense, because we expect the molecule
under higher tension to be more brittle \cite{Hyeon07JPHysCondMat}. These are the first indications that the fits using Eq.\ref{eq:2exp} do not provide meaningful parameters.
 
\subsection*{Structural basis of $f$-dependent switch in pathways:}
In order to provide a structural interpretation of the SMD nature of $f$-induced unfolding of src SH3, we followed the changes in several variables describing the conformations of SH3 as force is applied to residues 9 and 59 is varied in the SOP-SC simulations. Most of these are derived from measures assessing the extent to which  structures of various parts of the protein overlap  with the conformation in the native state. The structural overlap $\chi_{AB}$ for two parts of the protein $A$ and $B$ is the fraction of broken native contacts between $A$ and $B$\cite{Thirumalai1995},
\begin{equation}
	\chi_{AB}(\{\vect{r}\})=\frac{1}{M_{AB}}\sum_{\substack{i \in A\\ j \in B}}\Theta\left(\left||\vect{r}_i-\vect{r}_j|-|\vect{r}_i^0-\vect{r}_j^0|\right|-\Delta\right),
\end{equation}
where the summation is over the coarse-grained beads belonging to the parts $A$ and $B$, $M_{AB}$ is the number of contacts between $A$ and $B$ in the native state, $\Theta(x)$ is the Heaviside function, $\Delta=2\mathring{\mathrm{A}}$ is the tolerance in the definition of a contact, and $\vect{r}_{i,j}$ and $\vect{r}_{i,j}^0$, respectively, are the coordinates of the beads in a given conformation $\{\vect{r}\}$ and the native state. Two of the most relevant sets of contacts in the forced-rupture of SH3 are the ones between the N-terminal ($\beta 4$) and C-terminal ($\beta 5$) $\beta$-strands (Fig.\ref{fig:twoslopes}A), computed using  the structural overlap, $\chi_{\beta 4 \beta 5}$; and contacts between the RT-loop (residues 15 to 31) and the protein core (strands $\beta 1$ and $\beta 2$, residues 42 to 57) quantified by $\chi_\mathrm{RTL}$. When these structural elements unravel the structural overlap values  become close to unity, signaling the global unfolding of the SH3 domain.

Depending on $f$,  in some trajectories the RT-loop ruptures from the protein first ($\chi_\mathrm{RTL}$ sharply approaches 1), followed by the break between $\beta 4$ and $\beta 5$ strands ($\chi_{\beta 4 \beta 5} \approx 1$). In other trajectories, the order is opposite, with $\beta 4 \beta 5$ sheet melting first, without the RT-loop rupture (Fig.S1).The calculated the fraction, $P_{\Delta \mathrm{RTL}}$, of trajectories that unravel through rupture of the RT-loop pathway depends strongly on force, suggesting that these are the two major pathways responsible for the change in the slope of the [$\log k_\mathrm{u}(f),f$] plot (Fig.\ref{fig:snapshot}C). At low forces (15 pN) $P_{\Delta \mathrm{RTL}} \approx 0.8$ implying that $\approx 80\%$ of the trajectories unfold through the RT-loop pathway, and this fraction  decreases monotonically to $P_{\Delta \mathrm{RTL}}\approx 15\%$ at 45 pN. The movies in the Supplementary Information illustrate the two unfolding scenarios (See https://vimeo.com/150183198 for the RT-loop pathway and https://vimeo.com/150183352 for the $\beta 4 \beta 5$ pathway).


\subsection*{Effect of cysteine crosslinking:} In order to further illustrate that the slope change in Fig.\ref{fig:twoslopes} is due to the switching of the unfolding routes between the particular pathways discussed above, we created an \textit{in silico} mutant by adding a disulfide bond between the RT-loop and $\beta 2$ (mimicking a potential experiment with L24C/G54C mutant). In the crosslink mutant, the enhanced stability of the RT-loop to the protein core blocks the $\Delta \mathrm{RTL}$ unfolding pathway. We generated six 1500 ms unfolding trajectories at 15 pN and did not observe unfolding in any of them, thus obtaining an estimate for the upper bound of unfolding rate of $\approx  0.6$ s$^{-1}$ for this mutant. Comparing this unfolding rate to the rate at 15 pN for the wild-type (without the disulfide bridge) of 5.2 s$^{-1}$ shows that blocking of the pathway decreases the average unfolding rate at 15 pN. The mutant simulations with the disulfide bridge suggests that the RTL pathway plays a major role at low forces, and the unfolding through the $\beta 4 \beta 5$ pathways is much slower at low force. Furthermore, these simulations also show that rupture of the protein through the $\beta 4 \beta 5$ pathway occurs at a very slow rate at low forces even when the unfolding flux along the RTL pathway is muted.
Taken together these simulations explain the structural basis of rupture in the two major unfolding pathways.

\subsection*{Pathway switch occurs at a lower force in V61A mutant:}  To examine the effect of point mutations, we calculated $k_\mathrm{u}(f)$ as a function of $f$ for the V61A mutant. In the laser optical trapping (LOT) experiments, V61A mutant does not show upward curvature in the same force range, and the [$\log k_\mathrm{u}(f),f$] plot in that range is linear. However, the curvature can be seen at lower forces. In simulations, we observe the same qualitative change with respect to the wild-type upon mutation (Fig.\ref{fig:v61a}). If only data for forces above 15 pN is taken into account, the single exponential model becomes slightly more likely than the double exponential, but inclusion of the lower forces data shows double exponential, with pathway switching coming at a lower force than for WT. The fraction of trajectories going through the RT-loop pathway decreases compared to the wild-type (i.e. $P^{\mathrm{V61A}}_{\Delta\mathrm{RTL}}(f)<P^{\mathrm{WT}}_{\Delta\mathrm{RTL}}(f)$ for all $f$) (Fig. \ref{fig:v61a}C). The loss of upward curvature in the force range above 15 pN can be explained by the more prominent role of the $\beta 4 \beta 5$ pathway at low forces, leading to lesser degree of switching between the pathways. The V61A mutation is in the $\beta 5$ strand, making interactions between $\beta 4$ and $\beta 5$ weaker thus enabling the sheet to rupture more readily. Parenthetically we note that this is a remarkable result, considering that change in the SOP-SC force field is only minimal, which further illustrates that our model also captures the effect of point mutations.


\subsection*{Free energy profiles and transition states:}
Let us assume that the free energy landscape projected onto  extension as the reaction coordinate accurately captures the $f$-dependent unfolding kinetics. In this case, we expect the Bell model or its variation would hold, and $x$ (assumed to be $f$-independent) obtained from the fitting to that model would be the distance to the transition state with respect to the folded state $x^\ddag$. If the underlying free energy landscape were SMD it is still possible to formally construct a 1D free energy profile using experimental\cite{Hinczewski:2013kd} or simulation data. It is tempting to associate the distances in the projected 1D profiles with transition state locations with respect to the folded state, as is customarily done in analyzing SMFS data. Such an interpretation suggests that $x_L$ and $x_H$ should correspond to the distances to the two transition states in the two pathways, with $x^\ddag$ increasing with force in an apparent anti-Hammond behavior.
To assess if  this is realized, we constructed one-dimensional free energy profiles (of the WT protein) at forces 15, 30 and 45 pN to determine $x^\ddag$. It turns out, that $x^\ddag$ decreases rather than increases with force, demonstrating the normally expected Hammond behavior (Fig.\ref{fig:umbsam}), as force destabilizes the native state\cite{Hyeon:2006ii,Klimov:1999up}(see Discussion section).

We now demonstrate that $x_L$ and $x_H$ cannot be identified with transition state locations by calculating the committor probability, $P_{fold}$ \cite{Du98JCP}, the fraction of trajectories that reach the folded state before the unfolded state starting from $x_L$ or $x_H$. If $x_L$ and $x_H$ truly correspond to distances to transition states then $P_{fold} \approx 0.5$\cite{Du98JCP}, i.e., the transition state ensemble(TSE) should correspond to structures that have equal probability of reaching folded or unfolded state, starting from $x_L$ or $x_H$.  In sharp contrast to this expectation, the states with $x_L$ are visited hundreds of time before unfolding (see Fig.S3), which means $P_{fold}(x_L) \approx 1$. Thus, the usual interpretation of $x_L$ or $x_H$ ceases to have physical meaning, which is a consequence of the strong multidimensionality of the unfolding landscape of SH3. 

\subsection*{Force-dependent movement of the Transition State Ensemble:}
The results in Fig.S3 show that the extracted values of $x_L$ and $x_H$ cannot represent the transition state ensemble. Because the underlying reaction coordinates for the inherently SMD nature of folding landscapes are difficult to guess, the TSE can only be ascertained with a method that does not use a predetermined form of the reaction coordinate. We use the $P_{fold}$, based on the theory that the TSE should correspond to structures that have equal probability ($P_{fold} \approx 0.5$)of reaching the folded or unfolded state.  In order to determine the TSEs in our simulations, we picked the putative transition state structures from the saddle point of the 2D $(\chi,E)$ histogram of the unfolding trajectories ($0.66 < \chi < 0.73$; $92<E<105$ (kcal/mol) for $f=15$ pN and $0.59 < \chi < 0.73$; $82<E<106$ (kcal/mol) for $f=45$ pN), where $E$ is the total energy of the protein. We ran multiple trajectories from each of the candidate TS structures noting when the trajectory reaches the folded or the unfolded state first, in order to determine the $P_{fold}$. The set of structures with $0.4<P_{fold}<0.6$ is identified with the TSE. The $P_{fold}$ value for the whole ensemble is the total number of trajectories (starting from all the candidate structures) that reach the folded state first, divided by the total number of trajectories (or, the average of the individual $P_{fold}$ values). 

The TSE for 15 pN and 45 pN are given in Fig.\ref{fig:tse}. For both sets the $P_{fold}\approx 0.5 \pm 0.07$. The low-force TSE shows that the RT-loop is disconnected from the core ($\Delta\mathrm{RTL}$ state) and the 45 pN TSE has structures where the loop interacts with the core, but the contacts between N- and C- terminal $\beta$-strands are broken. The explicit TSE calculations confirm that the TSEs are similar to those found in unfolding trajectories with $\chi_{RTL}$ and $\chi_{\beta 4 \beta 5}$.

The experimental analysis of transition states of SH3 using mechanical $\phi$-values \cite{Guinn:2015gq} suggests that in the high force pathway the important residues are Phe-10 and Val-61 (which are in the $\beta 4$ and $\beta 5$), along with a core residue Leu-44. For the bulk (low/zero force) pathway, Phe-10, Ile-56 and Val-61 are also apparently important in TSE, as is the RT-loop residue Leu-24, which interacts with the protein core. Our simulation results, which provide a complete structural description of the TSEs, support the experimental interpretation, namely, loss of interaction between the RT-loop and the core at low forces and rupture of the $\beta 4 \beta 5$ sheet at high forces.

It is interesting to compute the mean extensions of the two major TSEs. The average distance between force application points for these structures is $x_{15}^{TSE} = 2.25 \pm 0.2$ nm for 15 pN and $x_{45}^{TSE} =2.7 \pm 0.3 $ nm for 45 pN, which (given the distances in the folded state of $x_{15}^{F}=1.96$ and $x_{45}^{F}=2.05$ nm) translates to the transition states of $x_{15}^\ddagger =0.29 \pm 0.18 $ and $x_{45}^\ddagger =0.65 \pm 0.33$ nm respectively. These values have no relation to $x_L$ and $x_H$, further underscoring the inadequacy of using Eq.\ref{eq:2exp} to interpret [$\log k_\mathrm{u}(f),f$] plots in SMD.

\section*{Discussion}

\subsection*{Hammond behavior:} Protein folding could be viewed using a chemical reaction framework. Just like in a chemical reaction, transitions occur from a minimum on a free energy landscape (corresponding to reactant or unfolded state) to another minimum (corresponding to a product representing the folded state, or an intermediate) by crossing a free energy barrier.  The top of the free energy barrier corresponds to a transition state.

Besides determining the structures of the unfolded and folded states, one of the main goals in protein folding is to identify the transition state ensemble, and characterize the extent of its heterogeneity. When viewed within the chemical reaction framework, the Hammond postulate provides a qualitative description of the structure of the transition state if it is unique. The Hammond postulate states \textit{``If two states, as, for example, a transition state and an unstable intermediate, occur consecutively during a reaction process and have nearly the same energy content, their interconversion will involve only a small reorganization of the molecular structure''}\cite{Hammond:1955cg}. A corollary of the Hammond postulate is that the TS structure likely resembles the least stable species in the folding reaction.

To apply the Hammond postulate to a protein free energy landscape, perturbed  by $f$, let us assume that at $f=f_m$ the states $F$ and $U$, with equal free energy, are separated by a transition state. Increasing $f$ will generally destabilize $F$, and  lower  the  free energy of $U$. According to Hammond's postulate, the transition state should be more similar to $F$ than $U$ as $f$ increases. If $f < f_m$, then the free energy of $F$ will be lower than $U$, and consequently the transition state will be more $U$-like. As a consequence of this argument, in unfolding induced by force, the transition state should move towards the state that is destabilized by $f$\cite{Klimov:1999up}, in accord with Hammond behavior. If the opposite were to happen  it could be an indication of anti-Hammond behavior.

In a one-dimensional energy landscape, the distance between a minimum and a barrier is reflected in the slope of the [$\log k(f), f$] plot ($m$ value for [$\log k([C]), [C]$]), which follows from the Arrhenius law and linear coupling in the free energy. Hammond behavior for the unfolding rate would mean movement of the transition state towards the folded state resulting in the decreasing of the slope of the [$\log k_\mathrm{u}(f), f$] plot with $f$. Hence, the temptation to refer to the opposite change of slope (i.e. increasing with $f$) as anti-Hammond behavior is natural. However, since the increase of the slope of the [$\log k_\mathrm{u}(f),f$] plot necessarily means, that the energy landscape is SMD, referring to movement of the transition state along a single reaction coordinate is not meaningful. Hence, the term ``anti-Hammond'' behavior in this case does not reflect the opposite of Hammond postulate in either the original formulation or according to the notion of the transition state. Moreover, even if the energy landscape is formally projected onto the reaction coordinate to which the parameter ($f$ or $[C]$) is coupled (which is possible even in the SMD case albeit without much physical sense), the movement of the transition state on this formal 1D landscape will still obey the Hammond postulate. Such a conclusion follows from a Taylor expansion of the first derivative of the perturbed (by $f$ or $[C]$) free-energy profile $F_f(x)=F_0(x)-fx$ around the barrier top ($x^\ddag$),
\begin{equation}
	\begin{split}
F^\prime_f(x)|_{x_f^\ddag}= &   \left(F_0(x)-fx\right)^\prime|_{x_f^\ddag}=F^\prime_0(x_0^\ddag+(x_f^\ddag-x_0^\ddag))-f=\\
= & F^\prime_0(x_0^\ddag)+F^{\prime\prime}_0(x_0^\ddag)(x_f^\ddag-x_0^\ddag)-f=0,
	\end{split}
	\label{eq:Taylorbarrier}
\end{equation}
where $F_0(x)$ and $x_0^\ddag$ are the free energy profile and transition state position at $f=0$. Since $F^\prime_0(x_0^\ddag)=0$ and $F^{\prime\prime}_0(x_0^\ddag)<0$, we find $x_f^\ddag-x_0^\ddag=f/F^{\prime\prime}(x_0^\ddag)<0$, or $x_f^\ddag<x_0^\ddag$, establishing that the transition state moves towards the native state, in accord with the Hammond behavior. Our conclusion holds for any perturbation $f$ which is linearly coupled to the energy function, and which monotonically destabilizes the folded state. Thus, we surmise that upward curvature in [$\log k_\mathrm{u}(f), f$] or [$\log k_\mathrm{u}([C]), [C]$] plots are not equivalent to anti-Hammond behavior. We note here, though the linear coupling of $f$ to the protein Hamiltonian is exact, the perturbation by denaturant is approximate, although the leading order in $[C]$ is linear. 

A similar conclusion, that is, a connection between upward curvature and multidimensionality, has been drawn analytically before, in the context of mechanochemistry of small molecules, based on the Taylor expansion of the Bell's model, similar to Eq.\ref{eq:Taylorbarrier}\cite{Konda:2011bw,Konda:2013fj}. 
In our work, we started from the most general description rather than from the solution of the Kramer's problem. 
The WMD conditions are similar to 1D assumptions when obtaining the Bell's model, but we do not make any assumptions about the barriers.
We also solved directly for the quantity we are interested in, i.e. sign of $\frac{d^2}{df^2}\log k_u(f)$, rather than movements of the transition state.
Connecting the latter to the curvature of the rate requires some additional steps, which might require more assumptions.

\subsection*{SMD in denaturant-induced unfolding:}
The criterion in Eq. \ref{d11} to assess if experiments can be be analyzed using a one-dimensional free energy profile applies to any external perturbation with a linear, additive contribution to the free energy. If  we consider the unfolding rate $k_\mathrm{u}([C])$ as a function of denaturant concentration $[C]$, a criteria analogous to Eq. \ref{d11} would hold if we assume that the energetic contribution due to $[C]$ is linear, proportional to a reaction coordinate related to the solvent-exposed surface area:
\begin{equation}
	U(x,\mb{r},[C]) = U_0(x,\mb{r}) + S(x) [C] 
\end{equation}
where $S(x)$ is the SASA-related monotone function of reaction coordinate $x$. Thus, for any perturbation ($f$ or $[C]$) coupling to the Hamiltonian, the theory and applications also hold when upward curvature in the [$\log k_\mathrm{u}([C]),[C]$] plot is observed.

Typically, the observed non-linearities in the [$\log k_\mathrm{u}([C]),[C]$] plots are analyzed using a double exponential fit, $k_\mathrm{u}([C]) = k_L^0\exp\left(m_L[C]\right) + k_H^0\exp\left(m_H[C]\right)$\cite{Wright:2003cs}
just like is done to analyze [$\log k_\mathrm{u}(f),f$] plots. Here, $m_L$ and $m_H$ are the analogues of $x_L$ and $x_H$ representing the unfolding $m$ values. It has been  shown for a protein with immunoglobulin fold \cite{Wright:2003cs} (see SI for the fits for several mutants of I27 using the double exponential model) and for monellin\cite{Aghera:2012dp}, that there is upward curvature in the [$\log k_\mathrm{u}([C]),[C]$] plots, which violates Eq. \ref{d11} implying that the underlying landscape in SMD. 
If [$\log k_\mathrm{u}([C]),[C]$] plots were linear then the unfolding $m$-value is likely to be proportional to the solvent accessible surface area in the transition state (even if the latter is heterogeneous), and the ensemble of conformations corresponding to  the $m$ value may be associated with the transition state ensemble ($P_{fold} \approx 0.5$). However,  for the [$\log k_\mathrm{u}([C]),[C]$] plots with upward curvature $m_L$ and $m_H$  may not  correspond to the SASAs of the respective transition states of the pathways just as we have shown that the extracted $x_L$ and $x_H$ should not be interpreted as TSE locations at low and high $f$, respectively. In addition, although the $k^\mathrm{H_2O}_L$ for the wild type is consistent with the expected value for $\beta$-sheet proteins with the I27 size, the $k^\mathrm{H_2O}_H\approx 10^{-4}k_L^0$ seems unphysical.  This observation combined with fairly high ratios of $\frac{m_H}{m_L}$ from a double exponential fit\cite{Wright:2003cs} (Table S1) suggests that although the double exponential model above fits the data, inferring the nature of the TSE requires entirely new set of experiments along the lines reported by Clarke and coworkers \cite{Wright:2003cs}. 



\subsection*{Pathway switch and propensity to aggregate:} In our previous work\cite{Zhuravlev:2014jk} we showed that an excited state $N^*$ in the spectrum of monomeric src SH3 domain has a propensity to aggregate. The structure of the $N^*$, which is remarkably close to the very lowly populated structure for $\it Fyn$ SH3 domain determined using using NMR\cite{Neudecker:2012kn}, has a ruptured interaction between $\beta 4$ and $\beta 5$. In other words, the value of $\chi_{\beta4 \beta5}$ is large. Interestingly, in our simulations unfolding of src SH3 domain occurs by weakening of these interactions at high forces (Figs. 3 and S1). Thus, the $N^*$ structures are dominant at high forces.  Because the probability of populating of $N^*$ is low at low forces ($N^*$ has 1-2\% probability of forming at $f=0$\cite{Zhuravlev:2014jk,Neudecker:2012kn}) it follows that SH3 aggregation is unlikely at low forces but can be promoted at high forces. Thus, SH3 domains have evolved to be aggregation resistant, and only under unusual external conditions they can form fibrils.

\subsection*{Prediction for a switch in the force unfolding of I27:} Based on our theory and simulations we can make a testable prediction for forced-unfolding of I27. Because there is upward curvature in the denaturant induced unfolding of I27\cite{Wright:2003cs}, we predict that a similar behavior should be observed for force-induced unfolding as well. In other words, there should be a switch in the pathway as the force used to unfold I27 is changed from a low to a high value. It is likely that this prediction has not been investigated because mechanical unfolding of I27 has so far been probed using only AFM\cite{Rief:1997ve}, where high forces are used. It would be most welcome to study the unfolding behavior of I27 using LOT experiments to test our prediction.

\section*{Conclusions}
We have proven that upward curvature in the unfolding rates as a function of a perturbation, which is linearly coupled to the energy function describing a protein in a solvent, implies that the underlying energy landscape is strongly multidimensional. The observation of upward curvature in the [$\log k_\mathrm{u}(f),f$] plots also implies that unfolding occurs by multiple pathways. In the case of $f$-induced unfolding of SH3 domain this implies that there is a continuous decrease in the flux of molecules that reach the unfolded state through the low force pathway as $f$ increases.  The numerical results using model two-dimensional free energy profiles allow us to conjecture that if a protein folds by parallel routes then the unfolding rate as a function of the linear perturbation must exhibit upward curvature. Only downward curvature in the [$\log k_\mathrm{u}(f),f$] plots can be interpreted using a single barrier one dimensional free energy profile with a moving transition state or one with two sequential barriers \cite{Merkel:1999Nature,Hyeon:2012JCP}.


Our study  leads to experimental proposals. For example, F\"{o}rster resonance energy transfer (FRET) experiments especially when combined with force would be most welcome to measure the flux through the two paths identified for src SH3 domain. Our simulations suggest that the FRET labels between the RT-loop and the protein core should capture the pathway switch, provided there is sufficient temporal resolution to observe the state with the RT-loop unfolded. A more direct way is to block the RT-loop pathway with a disulfide bridge between the RT-loop and the core, as we demonstrated using simulations, and assess if the unfolding rate decreases dramatically. Our work shows that the richness of data obtained in pulling experiments can only be fully explained by integrating theory and computations done under conditions that are used in these experiments.

\section*{Methods}
\subsection*{Force-dependent rates for SH3 domain using molecular simulations:} The 56 residue \textit{G.Gallus} src SH3 domain from Tyr kinase  consists of 5 $\beta$ strands (PDB 1SRL), which form $\beta$-sheets comprising the tertiary structure of the protein (see Fig.\ref{fig:twoslopes}A). Residues are numbered from 9 to 64. The details of the SOP-SC model are described elsewhere\cite{Zhuravlev:2014jk}. A constant force is applied to the N-terminal end (residue 9) and residue 59 (Fig.\ref{fig:twoslopes}A). 
We used Langevin dynamics, in the limit of high friction, in order to compute the $f$-dependent unfolding rates. We covered a range of forces from 12.5 pN to 45 pN generating between 50 -- 100 trajectories at each force. From these unfolding trajectories, we calculated the first time the protein unfolded ($x_{ee} > 5$ nm), thus  obtaining a set of times ${t_{ij}}$, for trajectory $j$ at force $f_i$.
We used umbrella sampling with weighted histogram analysis method\cite{WHAM} and low friction Langevin dynamics\cite{Honeycutt1992} to calculate free energy profiles.

\subsection*{Maximum Likelihood estimation:}
For the set of $M$ constant forces ${f_i}$, with $N_i$ measurements of the unfolding time at each force, assuming exponential distribution of unfolding times $P(t)=ke^{kt}$, where the rate $k$ depends on the force, the log-likelihood function is
\begin{equation}
	L=\log\prod\limits_{i}^M\prod\limits_{j}^{N_i} k(f_i)e^{-k(f_i)t_{ij}}=\sum\limits_{i}^M\sum\limits_{j}^{N_i}\log k(f_i)-k(f_i)t_{ij}.
\end{equation}
In the above equation $t_{ij}$ is the unfolding time measured in the $j$-th trajectory at force $f_i$. The exponential distribution allows us to take the sum over $j$ and use the average unfolding time $\tau_i=\sum\limits_j t_{ij}/N_i$ for each force, 
\begin{equation}
	L=\sum\limits_{i} N_i \left(\log k(f_i)-k(f_i)\tau_{i}\right).
	\label{eq:loglike}
\end{equation}
For each of the models (single- and double- exponential) the log-likelihood function $L$ was numerically maximized with the set of data $\{f_i,N_i; \tau_i\}$ (from simulations or experiment). The two maximal values of $L$ (for each model) were plugged into the Akaike information criterion\cite{Akaike:1974ih} to calculate relative likelihood of the models, i.e. the ratio of probabilities that the data is described by each of the models. The parameters that maximize $L$ are used for fitting the [$\log k_\mathrm{u}(f),f$] plots.

\subsection*{Akaike information criterion:}
The lower value of $AIC=2n-2L$ (where $L$ is the log-likelihood function and $n$ is the number of parameters in the model) indicates a more probable model, with the relative likelihood of models with $AIC_1$ and $AIC_2$ given by $P_2/P_1=\exp((AIC_1-AIC_2)/2)$. Thus, for the comparison of Bell's model ($L=L_1;n=2$) and double-exponential model ($L=L_2;n=4$), the double exponential is more probable by a factor of $P_2/P_1=\exp(L_2-L_1-2)$, where $L_1$ and $L_2$ are found by maximizing $L$ in Eq.\ref{eq:loglike}.


\section*{Acknowledgements} This work was supported by grants from the National Science Foundation (CHE 13-61946) and the National Institutes of Health (GM 089685)

\section*{Supplemental Info}
\subsection*{Fits for the titin I27 bulk experiment in denaturant:}
We performed maximum likelihood fitting for the data in [8] with single and double exponential models and compared the models using Akaike information criterion. The results are given in Table \ref{tbl:clarkefit}. Note the dramatic difference in the prefactors ($k_L^\mathrm{H_2O}$ and $k_H^\mathrm{H_2O}$), obtained using a double exponential fit, which is hard to explain. The difference in $m$-values, if they correspond to the Solvent Accessible Surface Area (SASA) in the transition state, does not appear to be meaningful. These observations suggest, just as in the case for force-induced unfolding, a de facto one dimensional fit does not yield physically meaningful results.

\subsection*{[$\log k_u(f), f$] plots for 2D landscapes:}
In order to better illustrate the connection between the curvature of the $[\log k, f]$ plot and existence of parallel pathways, we performed Brownian dynamics simulations of force-dependent rate of escape of a particle from the bound state for the landscapes given in Fig.1 of the main text. The resulting curves are given in panels (D,E,F). For each data point, we generated 8192 trajectories. The Fig.1A landscape is weakly multidimensional, so the $[\log k, f]$ plot does not exhibit upward curvature. For the landscape in Fig.1B, two parallel pathways exist, and flux through the states depend on $f$ as in experiments. The resulting curve has upward curvature. A double exponential fit is shown in panel (D). The landscape on Fig.1C gives rise to a more complex behavior.

\subsection*{Supplementary Movie 1.} An example of trajectory unfolding via the RT-loop pathway (https://vimeo.com/150183198).

\subsection*{Supplementary Movie 2.} An example of trajectory unfolding via the $\beta 4 \beta 5$ pathway (https://vimeo.com/150183352).

\bibliographystyle{pnas}


\begin{figure}
	\includegraphics[width=\textwidth]{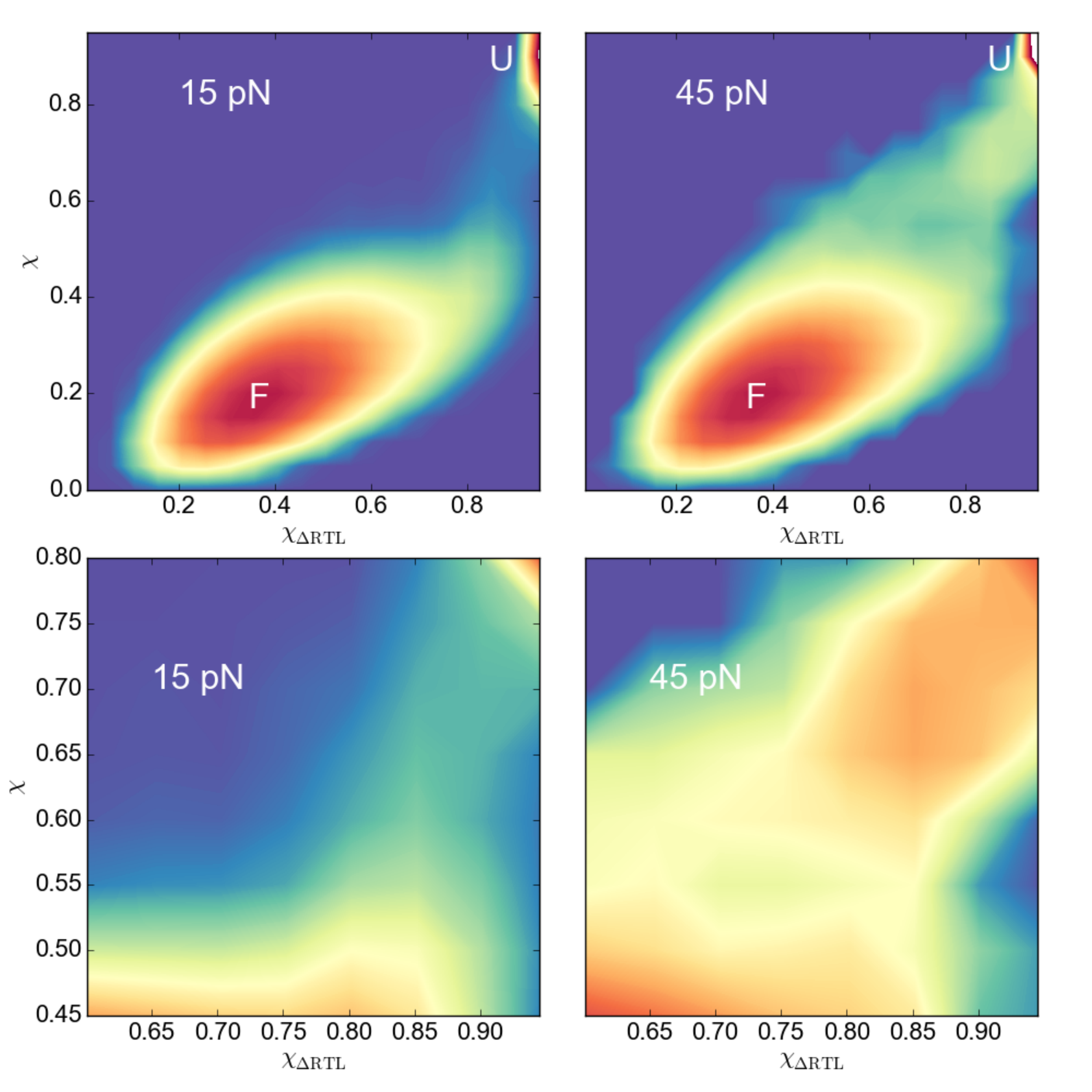}
\caption{Histograms of unfolding trajectories at 15 and 45 pN show the difference between unfolding pathways. The overlap parameter $\chi$ is the order parameter describing the global unfolding, while $\chi_{RTL}$ shows ruptured interactions between the RT-loop and the core. The lower panels are blowups of the route to the unfolded (U) state, whereas the upper panels show the profiles connecting F and U states (color scheme is also adapted to higher resolution). The forces are explicitly marked. The rupture process of the RT loop is dramatically different at 15 pN and 45 pN, demonstrating a switch in the unfolding pathway.}\label{fig:1545hist}
\end{figure}

\begin{figure}
	\includegraphics[width=\textwidth]{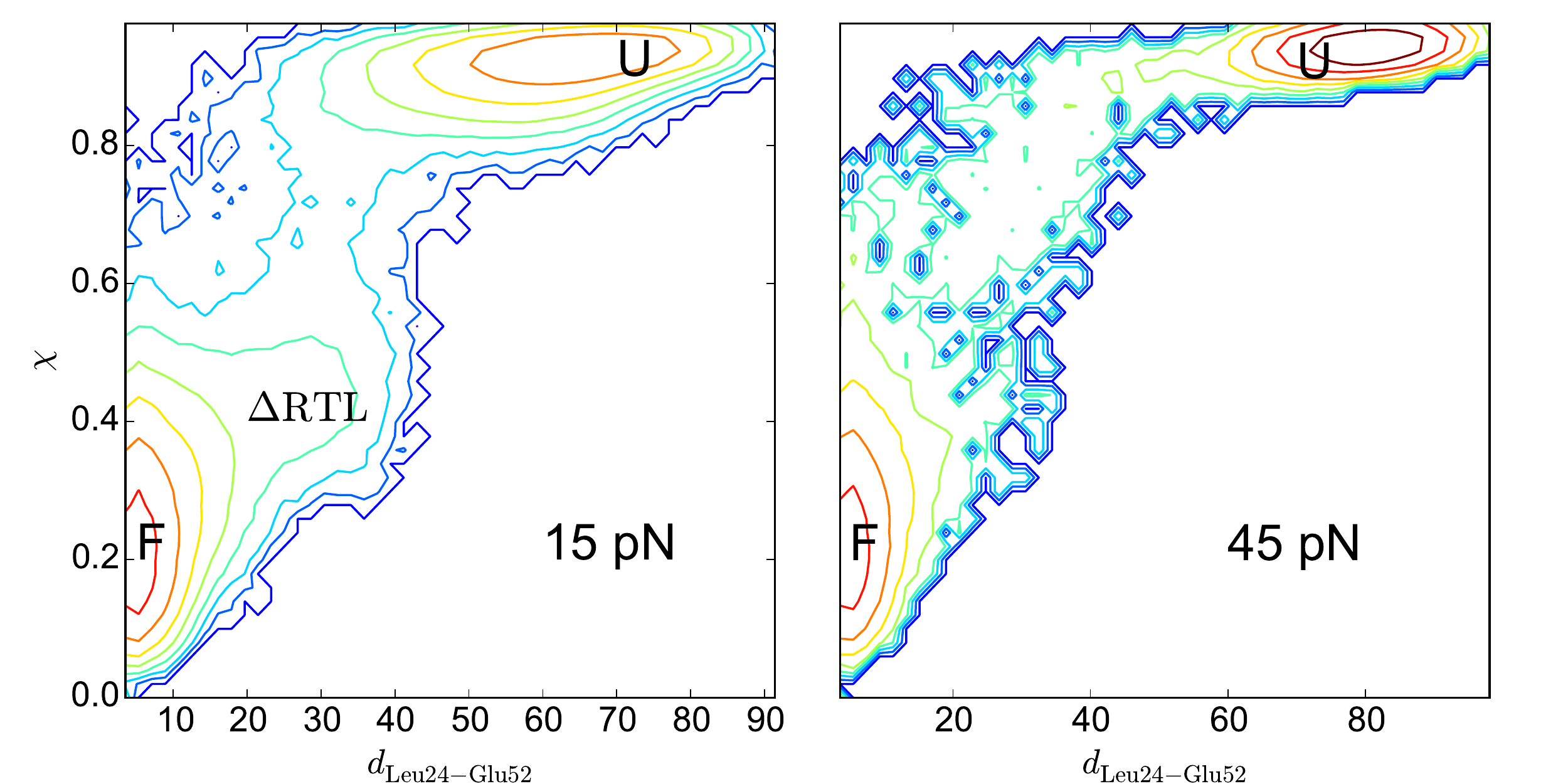}
\caption{Histograms of unfolding trajectories at 15 and 45 pN using a two-dimensional map with the global order parameter ($\chi$) and the distance between Leu24 and Glu52 side chains, which shows whether the RT-loop interacts with the protein core (Fig.\ref{fig:twoslopes}A). The profile offers an alternative view of the one in terms of $\chi$ and $\chi_{RTL}$ (Fig.\ref{fig:1545hist}).}\label{fig:L24Q52hist1545nofill}
\end{figure}

\begin{figure}
	\includegraphics[width=\textwidth]{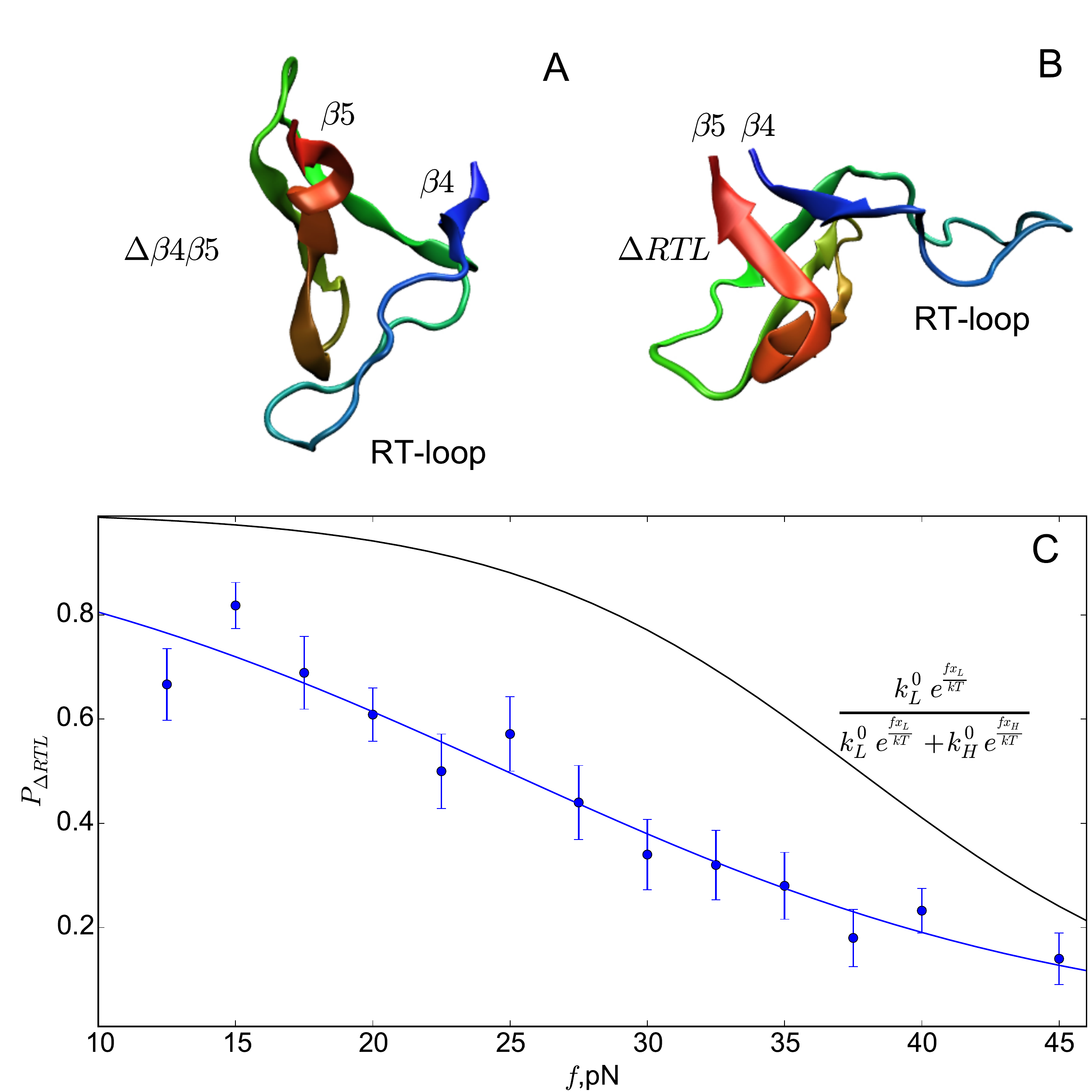}
   	\caption{(A) A snapshot of SH3 with the intact contacts between the RT-loop and the core but with ruptured $\beta 4$ and $\beta 5$ strands (state $\Delta \beta 4 \beta 5$). (B) A snapshot with the RT-loop peeled off, but $\beta 4 \beta 5$ intact (state $\Delta RTL$). (C) The flux of molecules through the $\Delta RTL$ state as a function of force. The blue symbols are obtained from simulations. The black line, which is the prediction  using the double exponential fit of the [$\log k_\mathrm{u}(f),f$] plot with Eq.\ref{eq:2exp}, differs from direct calculations based on simulations.
\label{fig:snapshot}}
\end{figure}

\begin{figure}
	\includegraphics[width=\textwidth]{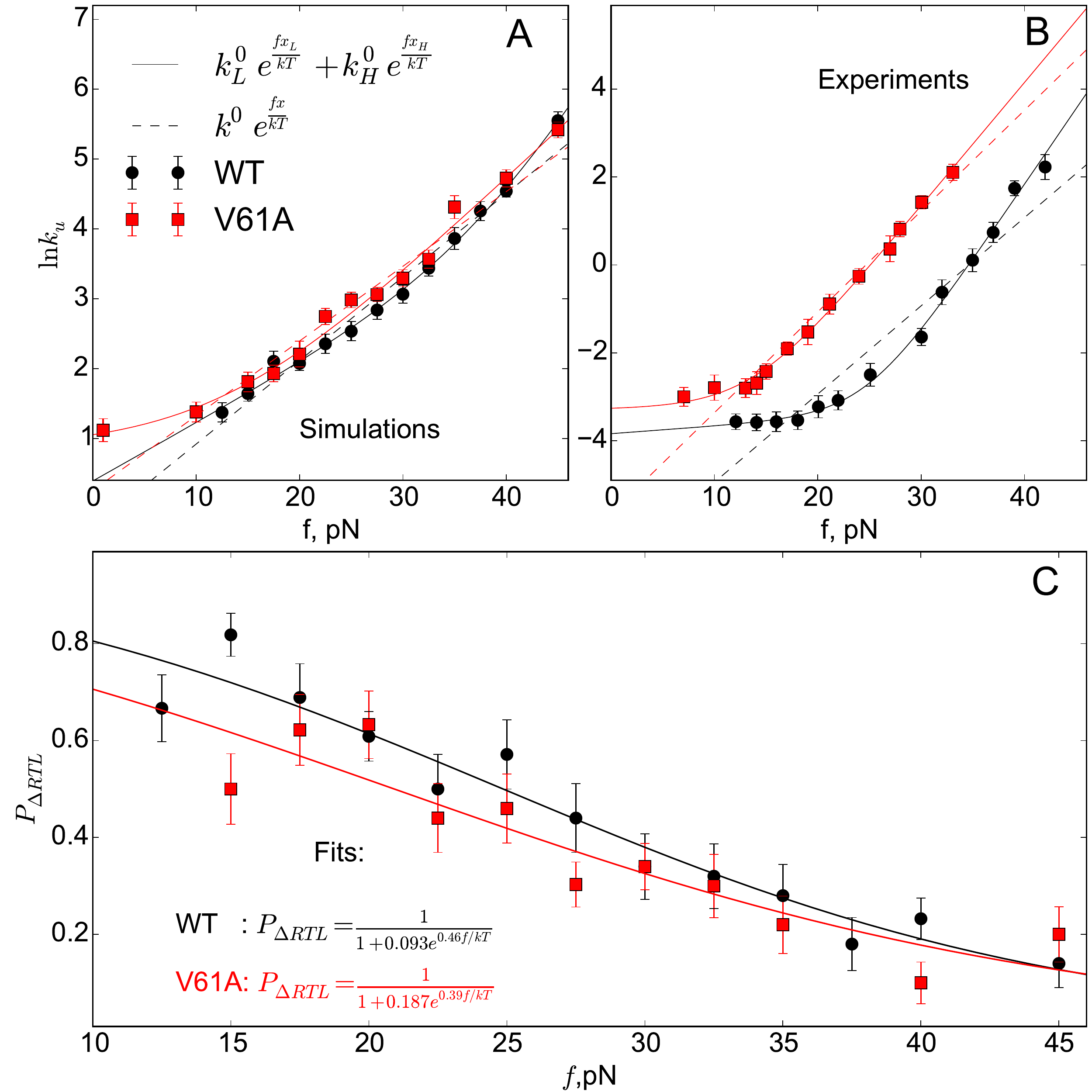}
   	\caption{(A) Unfolding rate for the V61A mutant obtained using simulations shows single exponential model to be more probable than the double exponential fit. (B) Same as (A) except the data are from experiments, (C) Fraction of trajectories that unfold through the low-force pathway as a function of $f$ obtained directly from simulations. Lines are the the fits to the function $(1+A\exp(B/kT))^{-1}$ with parameter values as indicated.
\label{fig:v61a}}
\end{figure}

\begin{figure}
	\includegraphics[width=\textwidth]{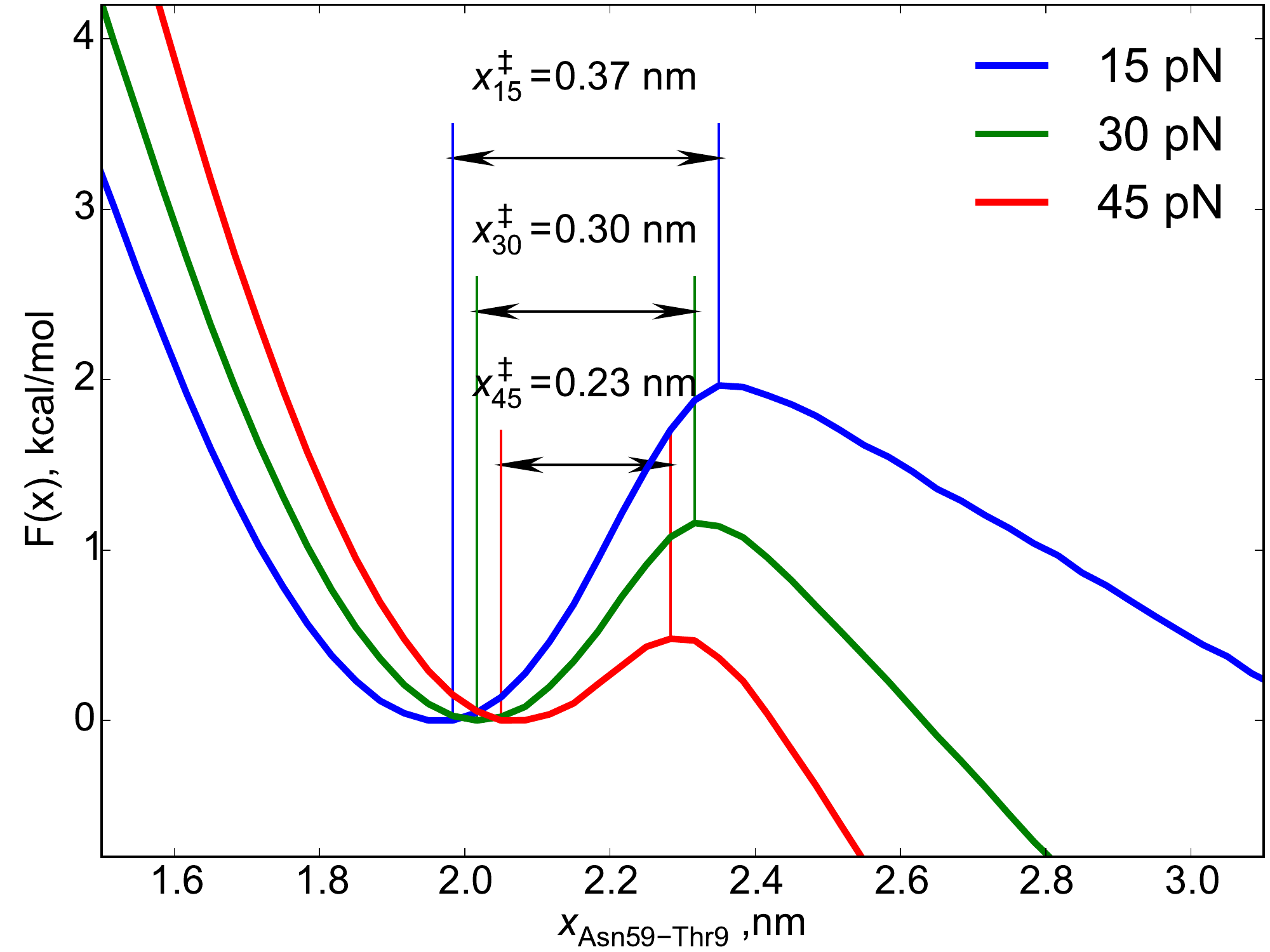}
   	\caption{One dimensional free energy profiles, $F(R)$, as a function of molecular extension, at three forces, calculated from simulations. The $F(R)$ profiles show that the ``distance to the transition state'' $x^\ddagger$ (the location of the top of the barrier with respect to the folded state) decreases as $f$ increases. Fits of the unfolding rate to the double exponential function produce the exact opposite behavior ($x_H>x_L$), which is an indication that the $x_L$ and $x_H$ extracted from  fitting the data to Eq.\ref{eq:2exp} do not coincide with $x^\ddag$.
\label{fig:umbsam}}
\end{figure}


\begin{figure}
\includegraphics[width=\textwidth]{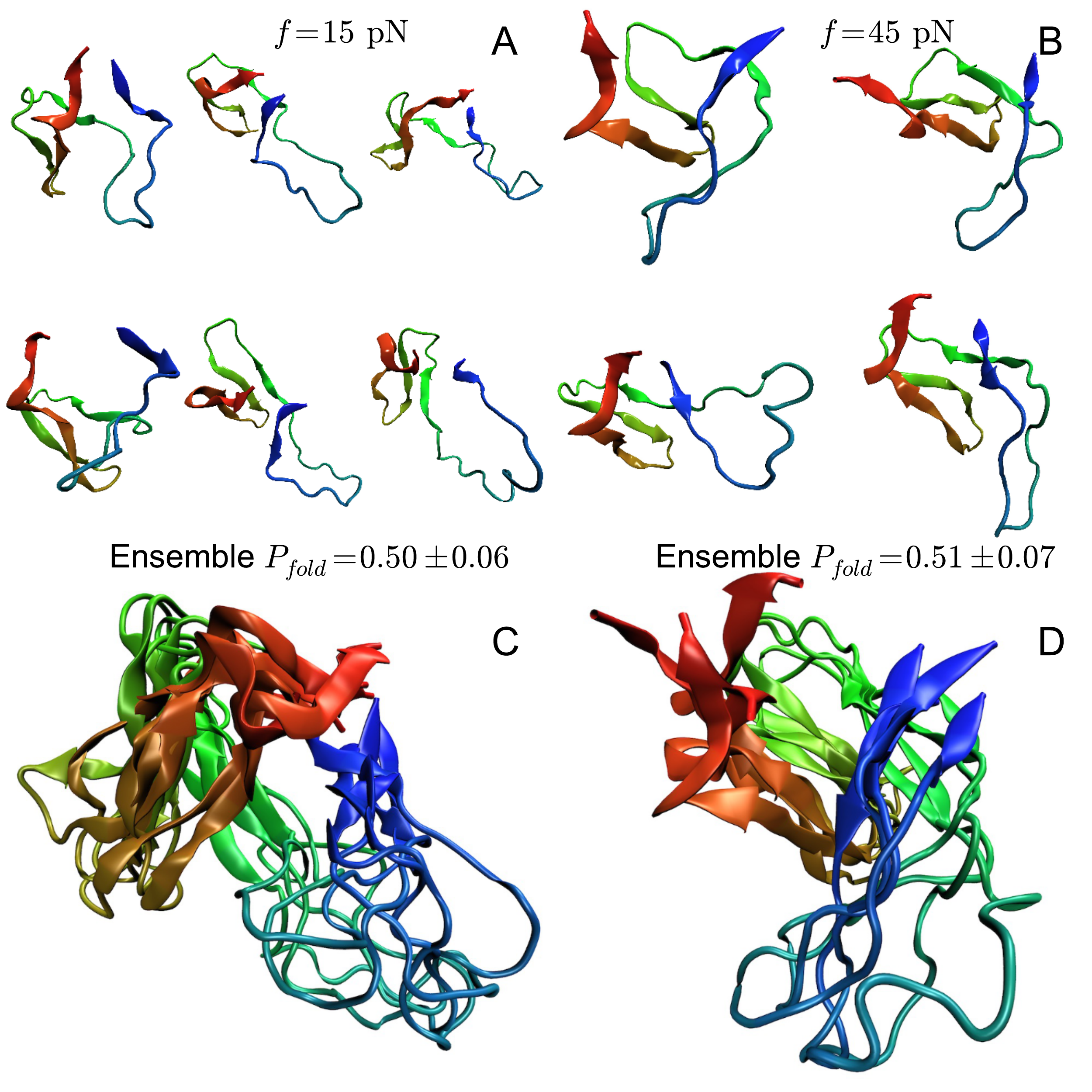}
\caption{Transition state ensembles structures at $f=15$ pN (A) and $f=45$ pN (B). (C) Overlay of the TSE at 15 pN and 45 pN (D). The TSEs were obtained using $P_{fold}$, which is independent of the reaction coordinate.}\label{fig:tse}
\end{figure}


\clearpage

\begin{table}\centering
  \caption{Comparison of fitting parameter sets for two different models for the rates obtained in our simulations and rates from the SMFS experiment\cite{Jagannathan:uw}. The definition of $P_2/P_1$ is given in Methods. The errors come out of log-likelihood covariance matrix with respect to parameters.}
  \label{tbl:fitparams}
  \begin{tabular}{llll||lll}
      \hline
  	 & \multicolumn{3}{c}{Simulations} & \multicolumn{3}{c}{Experiment}\\
  	\hline
      Name & Val. & Err. & Units & Val. & Err. & Units\\
      \hline
      $k^0_L$ & 1.5 & 0.4 & s$^{-1}$ & 2.0 & 1.1 & $10^{-2}$ s$^{-1}$ \\
      $k^0_H$ & 3.9 & 10.0 & $10^{-3}$ s$^{-1}$ & 6.1 & 6.4 & $10^{-6}$ s$^{-1}$\\
      $x_L$ & 0.40 & 0.1 &  nm & 0.08 & 0.1 & nm\\
      $x_H$ & 1.16 & 0.3 & nm & 1.42 & 0.12 & nm\\
	\hline
  	 & \multicolumn{3}{c}{$P_2/P_1 \sim 10^3$} & \multicolumn{3}{c}{$P_2/P_1 \sim 10^{17}$}\\

  \end{tabular}
\end{table}

\makeatletter \renewcommand{\fnum@table}
{\tablename~S\thetable}
\setcounter{table}{0}
\makeatother

\begin{table}
	\centering
  \caption{Fitting parameters for unfolding rates of titin I27 domain as a function of guanidinium chloride concentration for a set of single residue mutants and wild type [8]. Two types of fits were made: exponential fits $k_u = k^\mathrm{H_2O}\exp(m^\ddagger[C])$, and double exponential fits $k_u = k_L^\mathrm{H_2O}\exp(m_L[C]) + k_H^\mathrm{H_2O}\exp(m_H[C])$. $P_2/P_1$ shows relative likelihood of the double exponential model with respect to single exponential model, as assessed by Akaike information criterion. }
  \label{tbl:clarkefit}
  \begin{tabular}{lllll|ll|l}

	Name & \multicolumn{1}{c}{$k^\mathrm{H_2O}_L$} & \multicolumn{1}{c}{$k^\mathrm{H_2O}_H$} & $m_L$ & $m_H$ & \multicolumn{1}{c}{$k^{H_2O}$} & $m^\ddagger$ & $P_2/P_1$\\
	    \hline
	
		A75G & 6.3 & 4.7 & 0.28 & 1.35 & 2.2 & 0.54 & $\sim 10^{12}$\\
		
		C47A & 21.4 & 2.9 & 0.21 & 1.38 & 16.3 & 0.29 & $\sim 10^{19}$\\
		F21L & 14.8 & 0.005 & 0.41 & 2.40 & 6.6 & 0.58 & $\sim 10^{40}$\\
		G32A & 6.0 & 0.3 & 0.38 & 1.91 & 2.3 & 0.63 & $\sim 10^{42}$\\
		I23A & 3.2 & 4.3 & 0.26 & 1.36 & 0.8 & 0.64 & $\sim 10^{67}$\\
		I49V & 14.4 & 26.1 & 0.44 & 1.31 & 9.4 & 0.57 & $\sim 10^{8}$\\
		L36A & 44.4 & 6.0 & 0.33 & 1.60 & 25.0 & 0.49 & $\sim 10^{39}$\\
		L58A & 9.6 & 4.3 & 0.28 & 1.35 & 4.9 & 0.45 & $\sim 10^{8}$\\
		L60A & 43.8 & 4.1 & 0.29 & 1.58 & 26.0 & 0.43 & $\sim 10^{23}$\\
		L8A & 23.2 & 235.2 & 0.44 & 1.19 & 11.6 & 0.68 & $\sim 10^{14}$\\
		V30A & 9.0 & 4.3 & 0.30 & 1.35 & 4.4 & 0.47 & $\sim 10^{6}$\\
		V71A & 7.3 & 2.5 & 0.29 & 1.33 & 3.7 & 0.44 & $\sim 10^{0.7}$\\
		
		WT & 5.8 & 0.6 & 0.25 & 1.36 & 4.6 & 0.31 & $\sim 10^{6}$\\
		\hline

	Units & $10^{-4}\mathrm{s}^{-1}$ & $10^{-7}\mathrm{s}^{-1}$ & M$^{-1}$ & M$^{-1}$ & $10^{-4}\mathrm{s}^{-1}$ & M$^{-1}$ & \\

  \end{tabular}
\end{table}

\clearpage

\begin{table}
	\centering
  \caption{Fitting parameters for unfolding rates of SH3 domain as a function of force for wildtype and V61A mutant from simulations and experiment. Two types of fits were made: exponential fits $k_u = k^0\exp(fx/k_BT)$, and double exponential fits $k_u = k_L^0\exp(fx_L/k_BT) + k_H^0\exp(fx_H/k_BT)$. $P_2/P_1$ shows the relative likelihood of the double exponential model with respect to single exponential model, as assessed by Akaike information criterion.}
  \label{tbl:sh3fit}
  \begin{tabular}{lllll|ll|l}
	\hline
 	  Name & \multicolumn{1}{c}{$k^0_L$} & \multicolumn{1}{c}{$k^0_H$} & $x_L$ & $x_H$ & \multicolumn{1}{c}{$k^0$} & $x$ & $P_2/P_1$\\
 	    \hline
		Simulation WT & $1.47$ & $3.88\cdot 10^{-3}$ & 0.40 & 1.16 & $0.76$ & 0.58 & $10^{3}$\\
		Simulation V61A & $2.45$ & $4.20\cdot 10^{-1}$ & 0.02 & 0.67 & $1.29$ & 0.52 & $10^{5}$\\
		Experiment WT & $2.15\cdot 10^{-2}$ & $6.31\cdot 10^{-6}$ & 0.07 & 1.42 & $9.59\cdot 10^{-4}$ & 0.82 & $10^{17}$\\
		Experiment V61A & $3.83\cdot 10^{-2}$ & $8.46\cdot 10^{-4}$ & 0.0 & 1.15 & $3.50\cdot 10^{-3}$ & 0.94 & $10^{5}$\\
		\hline
 	Units & \multicolumn{1}{c}{s$^{-1}$} & \multicolumn{1}{c}{s$^{-1}$} & nm & nm & \multicolumn{1}{c}{s$^{-1}$} & nm & \multicolumn{1}{c}{1} \\
  \end{tabular}
\end{table}

\clearpage

\makeatletter
\makeatletter \renewcommand{\fnum@figure}
{\figurename~S\thefigure}
\setcounter{figure}{0}
\makeatother

\begin{figure}
	\centering
	\includegraphics[width=\textwidth]{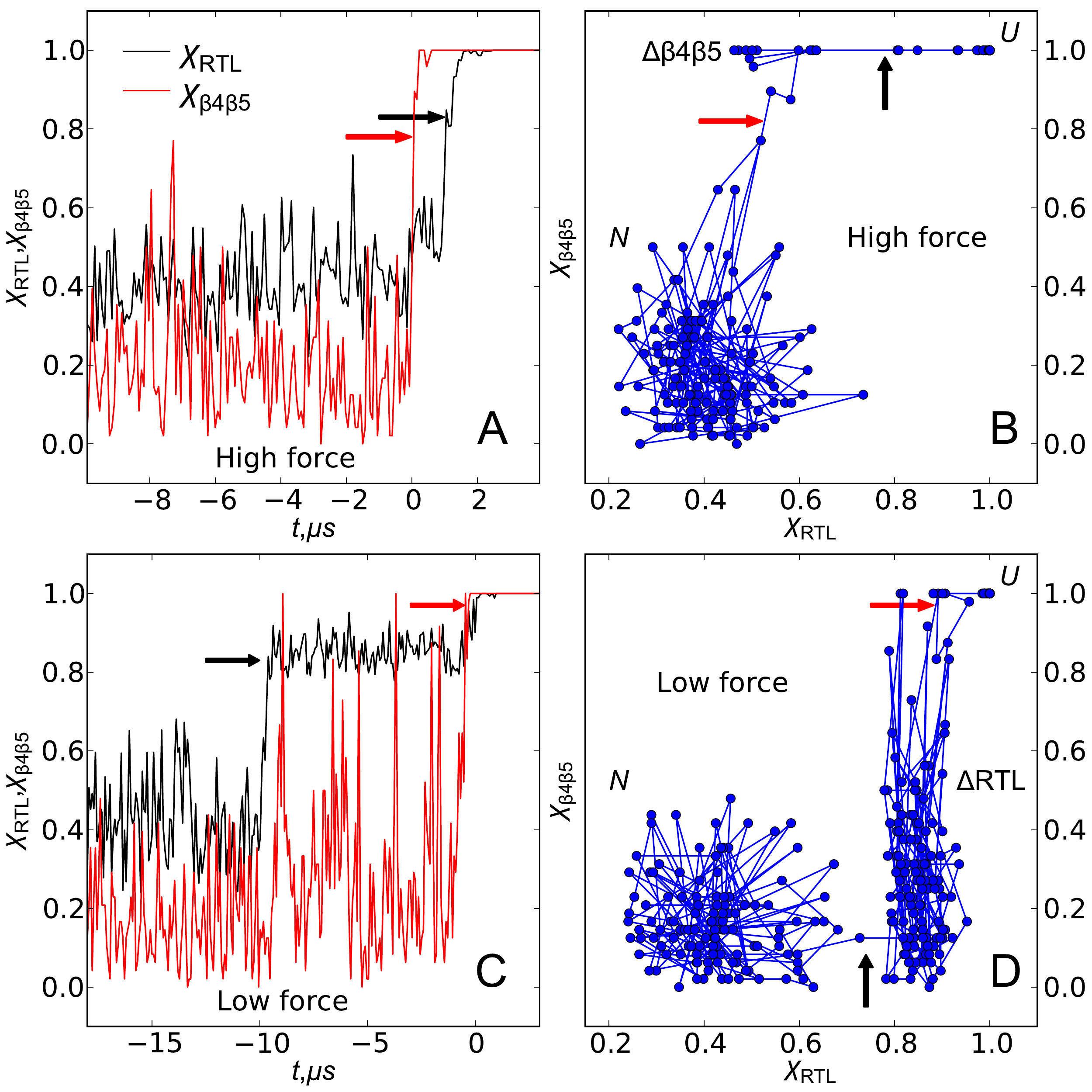}
   	\caption{Typical unfolding trajectories at high  (A,B) and low forces (C,D) in terms of fractions of ruptured native contacts ($\chi$) calculated for specific parts of the SH3 domain. $\chi_\mathrm{\beta 4 \beta 5}$ is for contacts between $\beta 4$ and $\beta 5$, while $\chi_\mathrm{RTL}$ is for contacts between the RT-loop and the protein core. At high forces, the $\beta 4 \beta 5$ contacts break first (red arrow in A,B), followed by the rest of the protein, including RT-loop (black arrow in A,B). There is a transient state $\Delta \mathrm{\beta 4 \beta 5}$ where the contacts between N-terminal and C-terminal $\beta$ strands are broken but the rest of the structure is intact. At low forces, the RT-loop is peeled from the core first (black arrow in C,D), while $\beta 4$-$\beta 5$ contacts are intact. The protein pauses in the $\Delta \mathrm{RTL}$ state for some length of time before the $\beta 4 \beta 5$ sheet melts (red arrow in C,D), followed by global unfolding.
\label{fig:twomechtraj}}
\end{figure}

\clearpage

\begin{figure}
\includegraphics[width=\textwidth]{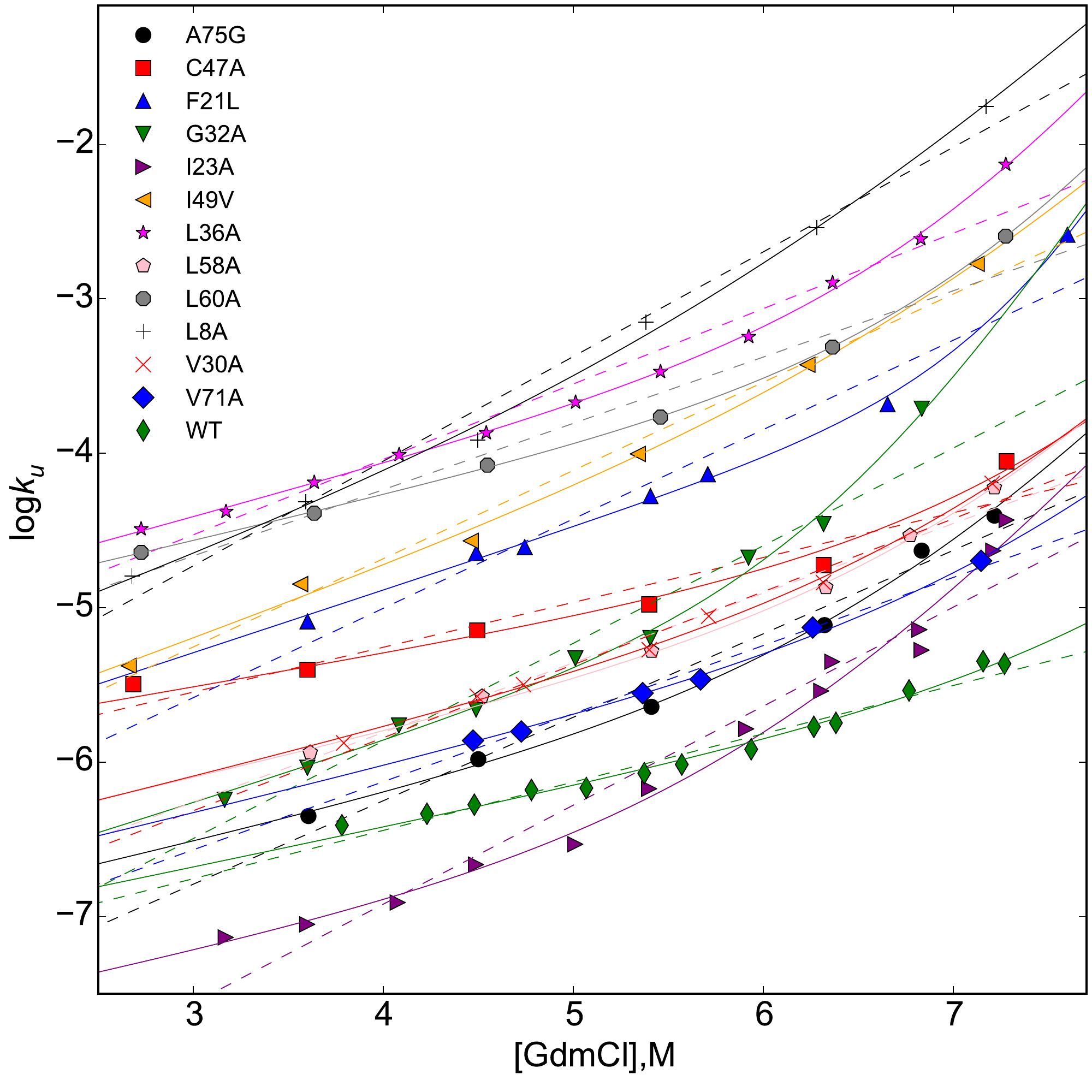}
\caption{Unfolding rates of titin I27 domain as a function of guanidinium chloride concentration for the wild type and various mutants [8]. Dashed lines show single exponential fits $k_u = k^\mathrm{H_2O}\exp(m^\ddagger[C])$, solid lines show double exponential fits $k_u = k_L^\mathrm{H_2O}\exp(m_L[C]) + k_H^\mathrm{H_2O}\exp(m_H[C])$. The extracted values of the parameters are in Table S\ref{tbl:clarkefit}.}\label{fig:clarke}
\end{figure}

\clearpage


\begin{figure}
	\centering
	\includegraphics[width=\textwidth]{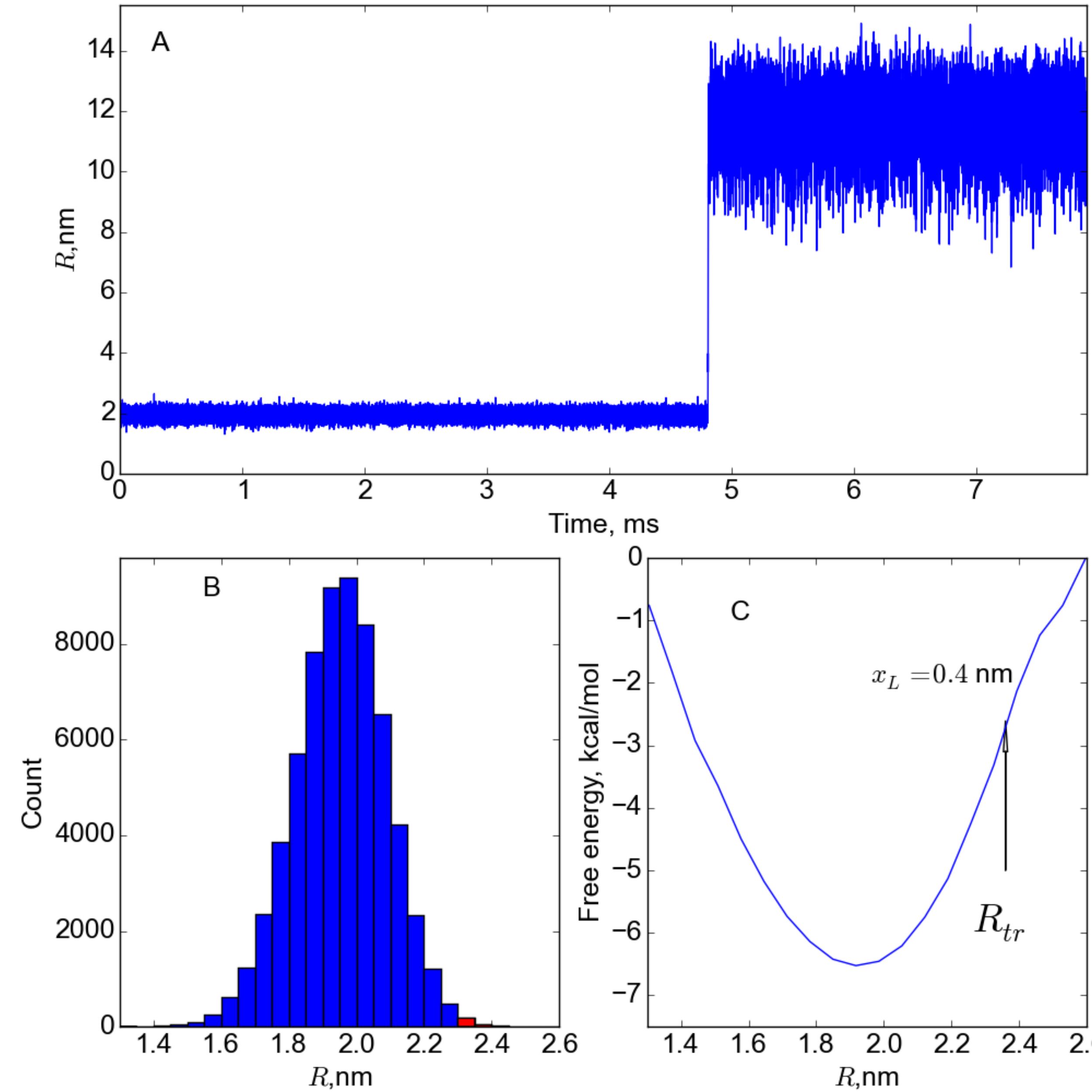}
   	\caption{(A) An example of a low force (15 pN) unfolding trajectory. (B) Histogram of the folded state calculated form the trajectory in (A). (C). Free energy profile $F(x)=-k_BT\log P(x)$. If $x_L$ is the distance to the transition state, then the transition state should be located at $\approx 2.34$ nm. The states corresponding to 2.34 nm are visited multiple times before the single unfolding event, but are always followed by the visit to the folded state. Therefore, $P_{fold}$ calculated from the ensemble of trajectories starting at $x_L$ is near 1. From the histogram in (B), we find that $x_L$ is visited 231 times before unfolding, which yields $P_{fold} \geq 0.996$.
\label{fig:pfold59L}}
\end{figure}

\end{document}